\documentclass[11pt]{article}

\usepackage{hyperref}
\hypersetup{colorlinks=false}

\usepackage{color}
\usepackage{latexsym}
\usepackage{amssymb}
\usepackage{amsmath}
\usepackage{graphicx}
\usepackage{subfig}
\usepackage{hhline}

\newtheorem{Remark}{Remark}[part]

\def \Sum{\displaystyle\sum}

\def \Frac{\displaystyle\frac}

\def \I{\mathbb{I}}
\def \N{\mathbb{N}}
\def \R{\mathbb{R}}

\def \F{\mathbb{F}}

\def \P{\mathbb{P}}

\def \S{\mathbb{S}}

\def\Pb{{\bf  P}}
\def\Eb{{\bf  E}}

\def \Ac{{\cal A}}

\def \Fc{{\cal F}}

\def \Lc{{\cal L}}
\def \Pc{{\cal P}}
\def \Qc{{\cal Q}}
\def \Mc{{\cal M}}

\def \Tc{{\cal T}}

\def \eps{\varepsilon}

\def\Dt#1{\Frac{\partial #1}{\partial t}}

\def\reff#1{{\rm(\ref{#1})}}

\def\beqs{\begin{eqnarray*}}
\def\enqs{\end{eqnarray*}}
\def\beq{\begin{eqnarray}}
\def\enq{\end{eqnarray}}

\begin{document}

\title{Optimal high frequency trading with limit and market orders}

%\author{Fabien Guilbaud\footnote{EXQIM, and LPMA, University Paris 7, fabien.guilbaud@exqim.com}~~~
%Huy\^en Pham\thanks{LPMA, University Paris 7, CREST-ENSAE, and Institut Universitaire de France, pham@math.jussieu.fr}
%}

\author{Fabien GUILBAUD
             \\\small  EXQIM  and 
             \\\small Laboratoire de Probabilit\'es et
             \\\small  Mod\`eles Al\'eatoires
             \\\small  CNRS, UMR 7599
             \\\small  Universit\'e Paris 7 Diderot
             \\\small  fabien.guilbaud@exqim.com
             \and
              Huy\^en PHAM
             \\\small  Laboratoire de Probabilit\'es et
             \\\small  Mod\`eles Al\'eatoires
             \\\small  CNRS, UMR 7599
             \\\small  Universit\'e Paris 7 Diderot
             \\\small  pham@math.jussieu.fr
             \\\small   CREST-ENSAE 
             \\\small  and Institut Universitaire de France
             }

\date{}

\maketitle

\begin{abstract}
We propose a framework for studying optimal market making policies in a limit order book (LOB). The bid-ask spread of the LOB is modelled by a Markov chain with finite values, multiple of the tick size, and subordinated by the Poisson process of the tick-time clock. We consider a small agent who continuously submits  limit buy/sell orders at best bid/ask quotes, and may also set limit orders at best bid (resp. ask) plus (resp. minus) a tick for getting the execution order priority, which is a crucial issue in high frequency trading. By trading with limit orders, the agent faces an execution risk  since her orders are executed only  when they meet counterpart market orders, which are modelled by Cox processes with intensities depending on the spread and on her limit prices.  By holding non-zero positions on the risky asset, the agent is also subject to the inventory risk related to price volatility. Then the agent can also choose to trade with market orders, and therefore get immediate execution, but at a least favorable price because she has to cross the bid-ask spread. 

%She can then post market orders at discrete trading times by crossing the spread, hence at a least favorable price, but for an immediate execution of her transaction.  

The objective of the market maker is to maximize her expected utility from re\-venue  over a short term horizon by a tradeoff between limit and market orders, while controlling her inventory position. This is formulated as a mixed regime switching re\-gular/impulse control problem that we characterize 
in terms of quasi-variational system  by dynamic programming methods.  In the case of a mean-variance criterion with martingale reference price or when the asset price follows a Levy process and  with exponential utility criterion, the dynamic programming system can be reduced to a system of 
simple equations involving only the inventory and spread variables. 

Calibration procedures are derived for estimating the transition matrix and intensity parameters for the spread and for Cox processes modelling the execution of limit orders. 
%We provide an explicit  backward splitting  scheme  for solving the problem, and show how it can be reduced to a system of 
%simple equations involving only the inventory and spread variables. 
Several computational tests are performed both on simulated and real data, and illustrate the impact and profit when considering execution priority in limit orders and market orders. 
\end{abstract}

\vspace{3mm}


\newpage

\section{Introduction}

Most of modern equity exchanges are organized as \textit{order driven} markets. In such type of markets, the price formation exclusively results from operating a \textit{limit order book} (LOB), an order crossing mechanism where \textit{limit orders} are accumulated while waiting to be matched with incoming \textit{market orders}. Any market participant is able to interact with the LOB by posting either market orders or limit orders\footnote{A market order of size $m$ is an order to buy (sell) $m$ units of the asset being traded at the lowest (highest) available price in the market, its execution is immediate; a limit order of size $\ell$ at price $q$ is an order to buy (sell) $\ell$ units of the asset being traded at the specified price $q$, its execution is uncertain and achieved only when it meets a counterpart market order. Given a security, the \textit{best bid} (resp. \textit{ask}) price is the highest (resp. lowest) price among limit orders to buy (resp. to sell) that are active in the LOB. The \textit{spread} is the difference, expressed in num\'eraire per share, of the best ask price and the best bid price, positive during the continuous trading session (see \cite{LOBsurvey}).}.

In this context, \textit{market making} is a class of strategies that consists in simultaneously posting limit orders to buy and sell during the continuous trading session. By doing so, market makers provide counterpart to any incoming market orders: suppose that an investor $A$ wants to sell one  share of a given security at time $t$ and that an investor $B$ wants to buy one share of this security at time $t'>t$; if both use market orders, the economic role of the market maker $C$ is to buy the stock as the counterpart of $A$ at time $t$, and carry until date $t'$ when she will sell the stock as a counterpart of $B$. The revenue that $C$ obtains for providing this service to final investors is the difference between the two quoted prices at ask (limit order to sell) and bid (limit order to buy), also called the market maker's spread. This role was traditionally fulfilled by specialist firms, but, due to widespread adoption of electronic trading systems, any market participant is now able to compete for providing liquidity. Moreover, as pointed out by empirical studies (e.g. \cite{men10},\cite{henjonmen10}) and in a recent review \cite{gri10} from AMF, the French regulator, this renewed competition among liquidity providers causes reduced effective market spreads, and therefore reduced indirect costs for final investors. 

Empirical studies (e.g. \cite{men10}) also described stylized features of market making strategies. First, market making is typically not directional, in the sense that it does not profit from security price going up or down. Second, market makers keep almost no overnight position, and are unwilling to hold any risky asset at the end of the trading day. Finally, they manage to maintain their \textit{inventory}, i.e. their position on the risky asset close to zero during the trading day, and often equilibrate their position on several distinct marketplaces, thanks to the use of high-frequency order sending algorithms. Estimations of total annual profit for this class of strategy over all U.S. equity market were around $10$ G\$ in 2009 \cite{gri10}.

Popular models of market making strategies were set up using a risk-reward approach. Three distinct sources of risk are usually identified: the inventory risk, the adverse selection risk and the execution risk. The \textit{inventory risk} \cite{avesto08} is comparable to the market risk, i.e. the risk of holding a long or short position on a risky asset. Moreover, due to the uncertain execution of limit orders, market makers only have partial control on their inventory, and therefore the inventory has a stochastic behavior. The \textit{adverse selection risk}, popular in economic and econo\-metric litterature, is the risk that market price unfavourably deviates, from the market maker point of view, after their quote was taken. This type of risk appears naturally in models where the market orders flow contains information about the fundamental asset value (e.g. \cite{fregrabau08}). Finally, the \textit{execution risk} is the risk that limit orders may not be executed, or be partially executed \cite{kuhstr10}. Indeed, given an incoming market order, the matching algorithm of LOB determines which limit orders are to be executed according to a price/time priority\footnote{A different type of LOB operates under {\it pro-rata} priority, e.g. for some futures on interest rates. In this paper, we do not consider this case and  focus on the main mechanism used in equity market.}, and this structure fundamentally impacts the dynamics of executions.

Some of these risks were studied in previous works. The seminal work \cite{avesto08} provided a framework to manage inventory risk in a stylized LOB. The market maker objective is to maximize the expected utility of her terminal profit, in the context of limit orders executions occurring at jump  times of Poisson processes. This model shows its efficiency to reduce inventory risk, measured via the variance of terminal wealth, against the symmetric strategy.  Several extensions and refinement of this setup can be found in recent litterature: \cite{gueferleh11} provides simplified solution to the backward optimization, an in-depth discussion of its characteristics and an application to the liquidation problem. In \cite{baylud11}, the authors develop a closely related model to solve a liquidation problem, and study continuous limit case. The paper  \cite{carjam11} provides a way to include more precise empirical features to this framework by embedding a hidden Markov model for high frequency dynamics of LOB. Some aspects of  the execution risk were also studied previously, mainly by considering the trade-off between passive and aggressive execution strategies. In \cite{kuhstr10}, the authors solve the Merton's portfolio optimization problem in the case where the investor can choose between market orders or limit orders; in \cite{ver11},  the possibility to use market orders in addition to limit orders is also taken into account, in the context of market making in the foreign exchange market. Yet the relation between execution risk and the microstructure of the LOB, and especially the price/time priority is, so far, poorly investigated.

In this paper we develop a new model to address these three sources of risk. The stock mid-price is driven by  a general Markov process, and we model the market spread as a discrete Markov chain that jumps according to a stochastic clock. Therefore, the spread takes discrete values in the price grid, multiple of the tick size. We allow the market maker to trade both via limit orders, which execution is uncertain, and via market orders, which execution is immediate but costly. The market maker can post limit orders at best quote or improve this quote by one tick. 
In this last case, she hopes to capture market order flow of agents  who are not yet ready to trade at the best bid/ask quote. 
Therefore, she faces a trade off between waiting to be executed at the current best price, or improve this best price, and then be more rapidly executed but at a less favorable price. We model the limit orders strategy as continuous controls, due to the fact that these orders can be updated at high frequency at no cost. On the contrary, we model the market orders strategy as impulse controls that can only occur at discrete dates. We also include fixed, per share or proportional fees or rebates coming with each execution. Execution processes, counting the number of executed limit orders, are modelled as Cox processes with intensity depending both on the market maker's controls and on the market spread. In this context, we optimize the expected utility from profit over a finite time horizon, by choosing optimally between limit and market orders, while controlling the inventory position. We study in detail  classical frameworks including mean-variance criterion and exponential utility criterion.

The outline of this paper is as follows. In section 2, we detail the model, and comment its features. We also provide direct calibration methods for all quantities involved in our model.  We formulate  in Section 3  the optimal market making control problem and derive the 
associated Hamilton-Jacobi-Bellman quasi variational inequality (HJBQVI) from dynamic programming principle. We show how one can reduce the 
number of state variables to the inventory and spread for the resolution to this QVI in two standard cases. 
%Section 4 is devoted to  the numerical scheme  for solving the HJBQVI and computing the optimal policy. We also examine several situations, %where we are able simplify this  algorithm by reducing the number of state variables to the inventory and spread. 
In the last section 4, we provide some numerical results and  empirical performance analysis.

\section{A market-making model}

\setcounter{equation}{0} \setcounter{Assumption}{0}
\setcounter{Theorem}{0} \setcounter{Proposition}{0}
\setcounter{Corollary}{0} \setcounter{Lemma}{0}
\setcounter{Definition}{0} \setcounter{Remark}{0}

\subsection{Mid price  and  spread process}

Let us fix a probability space $(\Omega,\Fc,\Pb)$ equipped with a filtration $\F$ $=$ $(\Fc_t)_{t\geq 0}$ satisfying the usual conditions.   It is assumed that all random variables and stochastic processes are defined on the stochastic basis $(\Omega,\Fc,\F,\Pb)$. 

The mid-price of the stock is an exogenous  Markov process $P$, with infinitesimal ge\-nerator $\Pc$ and state space $\P$. 
For example, $P$ is a L\'evy  process (e.g. an arithmetic Brownian motion),  or an exponential of L\'evy process (e.g. geometric Brownian motion). 
%\beq \label{dynP}
%dP_t &=& P_t (  \mu dt + \sigma dW_t),
%\enq
%with  constants $\mu$, $\sigma$ $>$ $0$, and $W$ is a standard one-dimensional Brownian motion.  
In the limit order book (LOB) for this stock, we consider a stochastic bid-ask spread resulting from the behaviour of market participants,   
taking  discrete values,  which are finite multiple of the tick size $\delta$ $>$ $0$, and jumping at random times. 
This is modelled as follows:  we first consider  the tick  time clock associated to a Poisson process $(N_t)_t$ with deterministic intensity $\lambda(t)$,  and representing the random times where the buy and sell orders of participants in the market affect the bid-ask spread. 
We next define  a discrete-time stationary Markov chain $(\hat S_n)_{n\in\N}$,  valued in the finite state space 
$\S$ $=$ $\delta \I_m$, $\I_m$ $:=$  $\{1,\ldots,m\}$, $m$ $\in$ $\N\setminus\{0\}$, with probability transition matrix $(\rho_{ij})_{1\leq i,j\leq M}$, i.e. 
$\Pb[\hat S_{n+1} = j\delta | \hat S_n = i\delta]$ $=$ $\rho_{ij}$, s.t. $\rho_{ii}$ $=$ $0$,  independent of $N$, 
and representing the random spread in tick time.  The spread process $(S_t)_t$ in calendar time 
is then defined as the time-change of $\hat S$ by $N$, i.e. 
\beq \label{dynS}
S_t &=& \hat S_{N_t}, \;\;\; t \geq 0. 
\enq
Hence, $(S_t)_t$ is a continuous time (inhomogeneous) Markov chain with intensity matrix $R(t)$ $=$ $(r_{ij}(t))_{1\leq i,j\leq m}$, where 
$r_{ij}(t)$ $=$ $\lambda(t)\rho_{ij}$ for $i\neq j$, and $r_{ii}(t)$ $=$ $-\sum_{j\neq i}r_{ij}(t)$. We assume that $S$ and $P$ are independent. 
 The best-bid and best-ask prices are defined by: $P_t^b$ $=$ $P_t - \frac{S_t}{2}$, $P_t^a$ $=$ $P_t+\frac{S_t}{2}$.

\subsection{Trading strategies in the limit order book}

We consider an agent (market maker), who trades the stock using either limit orders or market orders. 
She may submit limit buy (resp. sell) orders  specifying the quantity and the price she is willing to pay (resp. receive) per share, 
but will be executed only when an incoming sell (resp. buy) market order is matching her limit order. Otherwise, she can post market buy (resp. sell) orders for an immediate execution,  but, in this case obtain the opposite best quote, i.e. trades at the best-ask (resp. best bid) price, which is less favorable.

{\it Limit orders strategies.}   The agent  may submit  at any time  limit buy/sell orders  
at the current best bid/ask prices  (and then has to wait an incoming counterpart market order matching her  limit),  
but also control her own bid and ask price quotes by placing 
buy (resp. sell) orders at a marginal higher (resp. lower) price than the current best bid (resp. ask), i.e. at $P_t^{b_+}$ $:=$ $P_t^b+\delta$ 
(resp. $P_t^{a_-}$ $:=$ $P_t^a-\delta$). 
Such an alternative choice is used in practice by a market maker to capture market orders flow of undecided traders at the best quotes,  hence  to  get  priority in the order execution w.r.t. limit order at current best/ask quotes, and can be taken into account in our modelling with discrete spread of tick size $\delta$.

There is then a tradeoff between a larger performance for a quote at the current best bid (resp. ask) price, and a smaller  
performance for a quote at a higher bid price, but with faster execution. 
The submission and cancellation of 
limit orders are for free, as they provide liquidity to the market, and are thus stimulated.  
Actually, market makers receive some fixed rebate once their limit orders are executed.  The agent is assumed to be  
small in the sense that she does not influence the bid-ask spread. 
The limit order strategies are then modelled by a continuous time predictable control process: 
\beqs
\alpha_t^{make} &=& (Q_t^b,Q_t^a,L_t^b,L_t^a), \;\;\; t \geq 0,
\enqs
where $L$ $=$ $(L^b,L^a)$ valued in $[0,\bar\ell]^2$, $\bar\ell$ $>$ $0$, represents the size of the limit buy/sell order, and $Q$ $=$ $(Q^b,Q^a)$ represent the  possible  choices of  the bid/ask quotes either at best or at marginally improved prices, and valued in ${\cal Q}$ $=$ 
${\cal Q}^b\times{\cal Q}^a$, with $\Qc^b$ $=$ $\{ Bb, Bb_+\}$,  $\Qc^a$ $=$ $\{Ba,Ba_-\}$: 
%$\{ (B_{b},B_a), (q_{b},q_{a_-}),(q_{b_+},q_{a}),(q_{b_+},q_{a_-}) \}$:
\begin{itemize}
\item $Bb$:  best-bid quote,  and $Bb_+$: bid quote at best price plus the tick 
\item $Ba$: best-ask quote, and $Ba_-$: ask quote at best price minus the tick
%\item $q_{b_+a}$: bid quote at best price plus the tick, and best-ask quote 
%item $q_{b_+a_-}$: bid quote at best price plus the tick, and ask quote at best price minus the tick. 
\end{itemize}
Notice that when the spread is equal to one tick $\delta$,  a bid quote at best price plus the tick is actually equal to the best ask, and will then be considered as a buy market order. Similarly, an ask quote at best price minus the tick becomes a best bid, and is then viewed as a sell market order.  
In other words, the limit orders $Q_t$ $=$ $(Q_t^b,Q_t^a)$  take values in $\Qc(S_{t^-})$, where $\Qc(s)$ $=$ $\Qc^b\times\Qc^a$ when 
$s$ $>$ $\delta$,  $\Qc(s)$ $=$ $\{Bb\}\times\{Ba\}$ when $s$ $=$ $\delta$.  We shall denote by $\Qc_i^b$ $=$ $\Qc^b$ for $i$ $>$ $1$, and 
$\Qc_i^b$ $=$ $\{Bb\}$ for $i$ $=$ $1$, and similarly for $\Qc_i^a$ for $i$ $\in$ $\I_m$.

We  denote at any time $t$ by   $\pi^b(Q_t^b,P_t,S_t)$ and  $\pi^a(Q_t^a,P_t,S_t)$ the bid and ask  prices of the market maker, where the functions $\pi^b$ (resp. $\pi^a$) are defined on $\Qc^b\times\P\times\S$ (resp.  $\Qc^a\times\P\times\S$) by:
\beqs
\pi^b(q^b,p,s) & = &  \left\{ \begin{array}{cl}
				           p - \frac{s}{2}, & \mbox{ for } q^b \; = \; Bb \\
				            p - \frac{s}{2} + \delta & \mbox{ for } q^b \; = \; Bb_+.  
				          \end{array}
				\right. \\          
\pi^a(q^a,p,s) & = &   \left\{ \begin{array}{cl}
				          p + \frac{s}{2}, & \mbox{ for } q^a \; = \; Ba \\
				            p + \frac{s}{2} - \delta & \mbox{ for } q^a \; = \; Ba_-.
				          \end{array}
				\right.
\enqs
We shall denote by  $\pi^b_i(q^b,p)$ $=$ $\pi^b(q^b,p,s)$, $\pi^a_i(q^a,p)$ $=$ $\pi^a(q^a,p,s)$ for $s$ $=$ $i\delta$, $i$ $\in$ $\I_m$.  
 
\begin{Remark}
{\rm One can  take into account proportional rebates received by the market ma\-kers, by considering; 
$\pi^b(q^b,p,s)$ $=$ $(p-\frac{s}{2} + \delta 1_{q^b=Bb_+})(1-\rho)$,  $\pi^a(q^a,p,s)$ $=$ $(p+\frac{s}{2} - \delta 1_{q^a=Ba_-})(1+\rho)$, for some 
$\rho$ $\in$ $[0,1)$, or per share rebates with:  
$\pi^b(q^b,p,s)$ $=$ $p-\frac{s}{2} + \delta 1_{q^b=Bb_+} - \rho$,  $\pi^a(q^a,p,s)$ $=$ $p+\frac{s}{2} - \delta 1_{q^a=Ba_-} +\rho$, for some 
$\rho$ $>$ $0$.
}
\end{Remark} 
 
The limit orders of the agent are executed when they meet  incoming counterpart market orders. Let us then consider  the arrivals of market buy and market   sell  orders, which completely match  the limit sell and limit buy orders of the small agent, 
and modelled  by independent  Cox processes $N^a$ and $N^b$.  The intensity rate of $N_t^a$ is given by $\lambda^a(Q_t^a,S_t)$ where 
$\lambda^a$ is a deterministic function of the limit quote sell order, and of the spread, satisfying $\lambda^a(Ba,s)$ $<$ $\lambda^a(Ba_-,s)$. This 
natural condition  conveys the price/priority in the order execution in the sense that  an agent quoting a limit sell order at  ask price $P^{a_-}$ will be executed before traders at the higher ask price $P^a$, and hence receive more often market buy orders.  Typically, one would also expect that 
$\lambda^a$ is nonincreasing w.r.t. the spread, which means that the larger is the spread, the less often the market buy orders  arrive.   
Likewise, we assume that the intensity rate of $N_t^b$ is given by $\lambda^b(Q_t^b,S_t)$ where 
$\lambda^b$ is a deterministic function of the spread, and $\lambda^b(Bb,s)$ $<$ $\lambda^b(Bb_+,s)$.  We shall denote by 
$\lambda^a_i(q^a)$ $=$ $\lambda^a(q^a,s)$, $\lambda^b_i(q^b)$ $=$ $\lambda^b(q^b,s)$ for $s$ $=$ $i\delta$, $i$ $\in$ $\I_m$.

For a limit order strategy $\alpha^{make}$ $=$ $(Q^b,Q^a,L^b,L^a)$, the cash holdings $X$ and  the number of shares $Y$ hold by the agent (also called inventory)  follow the  dynamics
\beq
dY_t &=& L_t^b dN_t^b - L_t^a dN_t^a, \label{dynY} \\
dX_t &=& - \pi^b(Q_{t}^b,P_{t^-},S_{t^-})  L_t^b dN_t^b  + \pi^a(Q_t^a,P_{t^-},S_{t^-})  L_t^a dN_t^a.  \label{dynX} 
%\\
%&=&  - \; \big(P_{t^-} -\frac{S_{t^-}}{2} + \delta 1_{Q_t^b=Bb_+}\big)   L_t^b dN_t^b   \nonumber  \\
%& & \;\;\;\;\;  + \;  \big(P_{t^-} + \frac{S_{t^-}}{2} -  \delta 1_{Q_t^a=Ba_-}\big)   L_t^a dN_t^a,  \nonumber
\enq
%where $\rho$ $\in$ $(0,1)$ is a fixed proportional  rebate. 

 \vspace{2mm}
 
{\it Market order strategies.}  In addition to market making strategies, the investor may place market orders for an immediate execution reducing her inventory. The submissions of market orders, in contrast to limit orders,  take liquidity in  the market, and are usually subject to fees. We model market order strategies by an impulse control:
\beqs
\alpha^{take} &=& (\tau_n,\zeta_n)_{n\geq 0}, 
\enqs
where $(\tau_n)$ is an increasing sequence of stopping times representing the market order decision times of the investor, and $\zeta_n$, $n$ $\geq$ $1$,  are $\Fc_{\tau_n}$-measurable random variables valued in $[-\bar e,\bar e]$, $\bar e$ $>$ $0$, and giving the number of stocks purchased at the best-ask price if $\zeta_n$ $\geq$ $0$, or selled at the best-bid price if $\zeta_n$ $<$ $0$ at these times.   Again, we assumed that the agent is small so that her total market order will be executed immediately at the best bid or best ask price. In other words, we only consider a linear market impact,  which does not depend on the order size. When posting a market order strategy,  the cash  holdings and the inventory  jump at times $\tau_n$ by: 
\beq
Y_{\tau_n} &=& Y_{\tau_n^-} + \zeta_n, \label{sautY}\\
X_{\tau_n} &=&  X_{\tau_n^-} -  c(\zeta_n,P_{\tau_n},S_{\tau_n}) \label{sautX} 
%\\ &=& X_{\tau_n^-} - (\zeta_n + \eps |\zeta_n|) P_{\tau_n} - (|\zeta_n| + \eps\zeta_n) \frac{S_{\tau_n}}{2}, \nonumber
\enq
where 
\beqs
c(e,p,s) &=& 
%e(p+\frac{s}{2})1_{e>0} +  e(p-\frac{s}{2})1_{e<0} \; = \; 
 ep + |e|\frac{s}{2} + \eps
\enqs 
represents the (algebraic) cost function indicating the amount to be paid immediately when passing  a market order of size $e$, given the mid price 
$p$, a spread $s$, and a fixed fee $\eps$ $>$ $0$. 
We shall denote by $c_i(e,p)$ $=$ $c(e,p,s)$ for $s$ $=$ $i\delta$, $i$ $\in$ $\I_m$. 

\begin{Remark}
{\rm One can also include proportional fees $\rho$ $\in$ $[0,1)$ paid at each market order trading by considering a cost function in the form: 
%\beqs
%c(e,p,s)  e(1+\eps)(p+\frac{s}{2})1_{e>0} +  e(1-\eps)(p-\frac{s}{2})1_{e<0} \; = \; 
$c(e,p,s)$ $=$ $(e+\eps|e|)p + (|e|+\rho e)\frac{s}{2}+\eps$, or fixed fees per share with 
$c(e,p,s)$ $=$ $ep+|e|(\frac{s}{2}+\rho)+\eps$. 
%\enqs 
 }
\end{Remark}

 \vspace{2mm}

In the sequel, we shall denote by $\Ac$ the set of all limit/market order trading strategies $\alpha$ $=$ $(\alpha^{make},\alpha^{take})$.  
 
\subsection{Parameters estimation}
 
In most order-driven markets, available data are made up of \textit{Level 1 data} that contain transaction prices and quantities at  best quotes,  
and of  \textit{Level 2 data}  containing  the volume updates for the liquidity offered at the $L$  first  order book slices  ($L$ usually ranges from 5 to 10). In this section, we propose some direct  methods for  estimating  the intensity  of the spread Markov chain, and of the execution point processes, based only on the observation of   \textit{Level 1 data}. This has the advantage of low computational cost, since we do not have to deal with the whole volume of \textit{Level 2 data}. Yet, we mention some recent  work  on parameters estimation  from the whole order book data 
\cite{constotal10}, but involving  heavier computations based on integral transforms.

\vspace{2mm}

\noindent  {\bf  \em{Estimation of spread parameters}.}  Assuming that the spread $S$ is observable,  
let us define the jump times of the spread process:
\beqs
\theta_0 \; = \; 0, \;\;\;\;\;  \theta_{n+1} &=& \inf \left\lbrace t > \theta_n \, :\, S_t \neq S_{t-}\right\rbrace \, \, , \, \forall  n \geq 1. 
\enqs
From these observable quantities, one can reconstruct the processes:
\beqs
N_t &=& \# \left\lbrace \theta_j >0 \, :\, \theta_j \leq t  \right\rbrace \, \, , \,  t \geq 0, \\
\hat{S}_n &=& S_{\theta_n} \, \, , \, \;  n \geq 0.
\enqs
Then, our goal is to estimate the deterministic intensity of the Poisson process $(N_t)_t$, and the transition matrix of the Markov chain $(\hat S_n)_n$ from a path realization with high frequency data of the tick-time clock and spread in tick time  over a finite trading time horizon $T$, typically of one day.  
From  the  observations of $K$ samples of $\hat S_n$, $n$ $=$ $1,\ldots,K$,  and since the Markov chain 
$(\hat S_n)$ is stationary,  we have a 
consistent estimator (when $K$ goes to infinity)  for the transition probability $\rho_{ij}$ $:=$  $\Pb[\hat S_{n+1} = j\delta | \hat S_n = i\delta]$ $=$ 
$\Pb[ (\hat S_{n+1},\hat S_n) = (j\delta,i\delta)]/\Pb[\hat S_n=i\delta]$  given by:    
\beq \label{estimrho}
\hat\rho_{ij} & = & \frac{  \Sum_{n=1}^K 1_{\lbrace (\hat{S}_{n+1},\hat S_n) = (j\delta,i\delta) \rbrace} } { \Sum_{n=1}^K 1_{\lbrace \hat{S}_{n} = i\delta \rbrace}}
\enq
For the estimation of the deterministic intensity function $\lambda(t)$ of the (non)homogeneous Poisson process $(N_t)$,  we shall assume 
in a first approximation a simple natural parametric form.  
%by considering that  $\lambda$ is piecewise constant with known jump times.  
For example, we may assume that  $\lambda$ is constant over a trading day, and more realistically for taking into account intra-day seasonality effects, we consider that the tick time clock intensity jumps e.g. every hour of a trading day. We then assume that $\lambda$ is in the form: 
\beqs
\lambda (t) &=&  \sum \lambda_k 1_{\lbrace t_{k}\leq t < t_{k+1} \rbrace}
\enqs
where $(t_k)_k$ is a fixed and known increasing finite sequence of $\R_+$ with $t_0=0$, 
and $(\lambda_k)_k$ is an unknown  finite sequence of $(0,\infty)$.  In other words, the intensity is constant 
equal to $\lambda_k$ over each period  $[t_k,t_{k+1}]$,  and by assuming that the interval length $t_{k+1}-t_k$ is large w.r.t. the intensity $\lambda_k$ (which is the case for high frequency data), 
we have a consistent estimator of  $\lambda_k$, for all $k$, and then of $\lambda(t)$ given by: 
\beq \label{estimlambda}
\hat{\lambda}_k &=&  \frac{N_{t_{k+1}}-N_{t_{k} } } {t_{k+1} - t_{k}}. 
\enq

We performed these two estimation procedures \reff{estimrho} and \reff{estimlambda} on SOGN.PA stock on April 18, 2011 between 9:30 and 16:30 in Paris local time. We used tick-by-tick level 1 data provided by Quanthouse, and fed via a OneTick Timeseries database.

In table \ref{transitionmatrix},  we display the estimated transition matrix: first row and column indicate the spread value $s=i\delta$ and the the cell $ij$ shows $\hat{\rho}_{ij}$. 
For this stock and at this date, the tick size was $\delta=0.005 $ euros, and we restricted our analysis to the first $6$ values of the spread $(m=6)$ due to the small number of data outside this range. 
%\marginpar{cette estimation est valable seulement a cette date!}

\begin{table}[h!]
\footnotesize
\begin{center}
\begin{tabular}{|l|llllll|}
\hline	 spread &	0.005	&	0.01	&	0.015	&	0.02	&	0.025	&	0.03	\\
\hline 0.005	&	0	&	0.410	&	0.220	&	0.160	&	0.142	&	0.065	\\
0.01	&	0.201	&	0	&	0.435	&	0.192	&	0.103	&	0.067	\\
0.015	&	0.113	&	0.221	&	0	&	0.4582	&	0.147	&	0.059	\\
0.02	&	0.070	&	0.085	&	0.275	&	0	&	0.465	&	0.102	\\
0.025	&	0.068	&	0.049	&	0.073	&	0.363	&	0	&	0.446	\\
0.03	&	0.077	&	0.057	&	0.059	&	0.112	&	0.692	&	0	\\
\hline
\end{tabular}
\end{center}
\caption{Estimation of the transition matrix $(\rho_{ij})$ for  the underlying spread of  the stock SOGN.PA on  April 18, 2011.  }
\label{transitionmatrix}
\end{table}

\begin{figure}[h!] 
\centering
\includegraphics[width=0.9\textwidth]{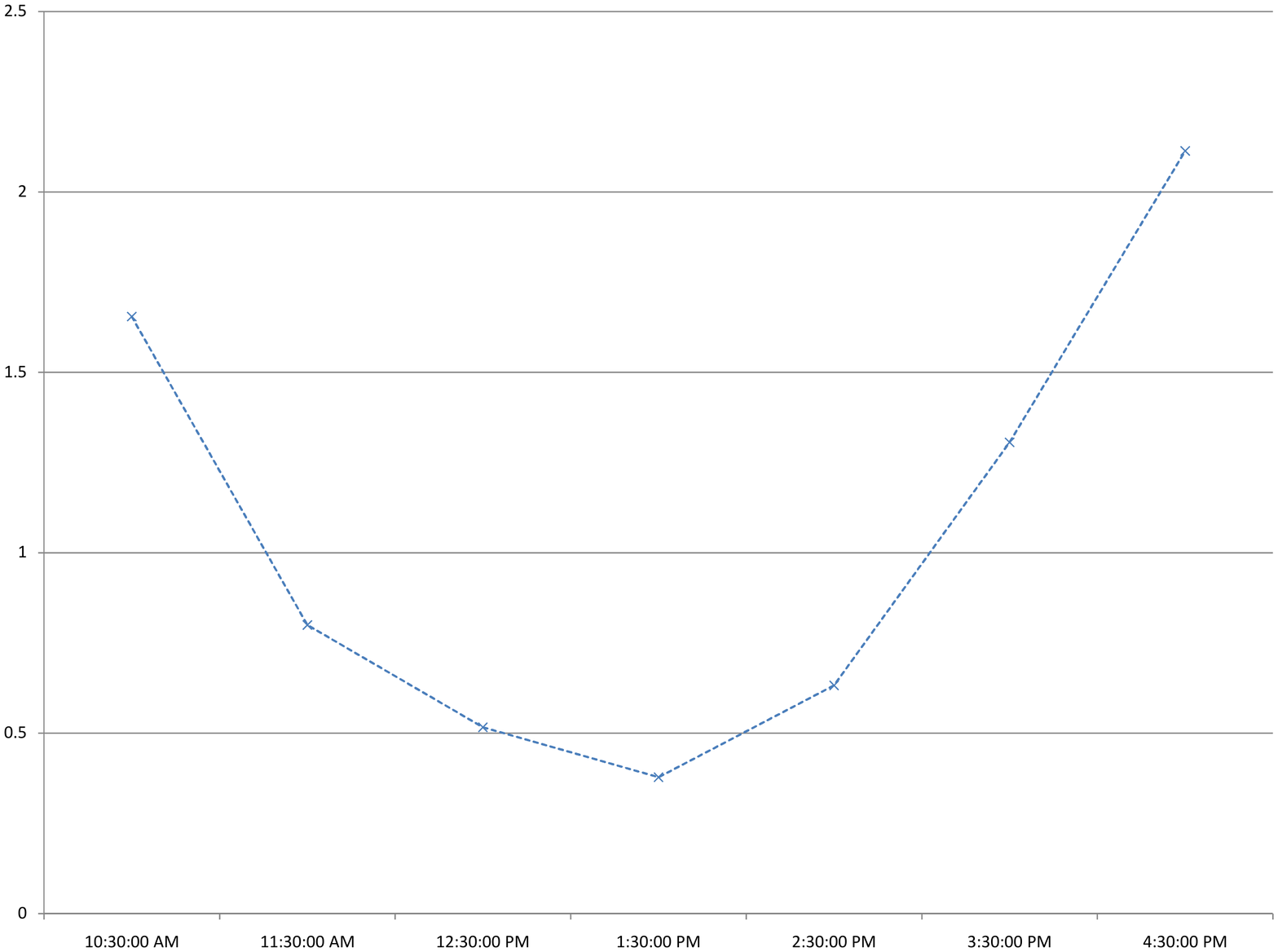}
\caption{Plot of tick time clock intensity estimate for the stock SOGN.PA on April 18, 2011 expressed in second$^{-1}$ (affine interpolation). }
\label{clockintensityplot}
\end{figure}

In table \ref{clockintensity},  we display  the estimate for the tick time clock intensity $\lambda (t)$ by assuming that it is piecewise constant and jumps only at 10:30, 11:30, ..., 16:30 in Paris local time. In figure \ref{clockintensityplot}, we plot the tick time clock intensity by using an affine interpolation, and observed a typical U-pattern.   Hence, a  further step for the estimation of the intensity  could be to specify a parametric form for the intensity function fitting $U$ pattern, e.g. parabolic functions in time,  and then use a maximum likelihood method for estimating the parameters.

\begin{table}[h!]
\footnotesize
\begin{center}
\begin{tabular}{|l|l|l|l|l|l|l|l|}
\hline 
Time	&	10:30 &	11:30	&	12:30	&	13:30	&	14:30&	15:30	&	16:30\\
\hline Clock Intensity	$\lambda(t)$ &	1.654	&	0.799	&	0.516	&	0.377	&	0.632	&	1.305	&	2.113	\\
\hline
\end{tabular}
\end{center}
\caption{Estimation of the tick time clock (hourly basis) for the stock SOGN.PA on April 18, 2011. Tick time clock intensity $\lambda(t)$ is expressed in second$^{-1}$.}
\label{clockintensity}
\end{table}

\vspace{2mm}

\noindent {\bf \em{Estimation of execution parameters}.} 
When performing a limit order strategy $\alpha^{make}$, we suppose that the market maker permanently monitors her execution point processes $N^a$ and $N^b$, representing respectively 
the number of arrivals of market buy and sell orders matching the limit orders  for quote ask $Q^a$ and quote bid $Q^b$.  
We also assume that there is no latency so that the observation of the execution processes is not noisy. Therefore, observable variables include the quintuplet:
\beqs
(N^a_t,N^b_t,Q^a_t,Q^b_t,S_t) \in \R^+ \times \R^+ \times \Qc^a \times \Qc^b\times\S \, , \, t \in [0,T]
\enqs
Moreover, since $N^a$ and $N^b$ are assumed to be independent, and both sides of the order book can be estimated using the same procedure,  we shall  focus on the estimation for 
the intensity function $\lambda^b(q^b,s)$, $q^b$ $\in$ $\Qc^b$ $=$ $\{Bb,Bb_+\}$, $s$ $\in$ $\S$ $=$ $\delta\I_m$,  of the Cox process $N^b$. 
%Therefore the observable variables of interest are:
%\beqs
%(N^b_t,Q^b_t,S_t) \in \R^+\times \Qc^b\times\S \, , \, t \in [0,T]
%\enqs

The estimation procedure for $\lambda^b(q^b,s)$ basically matchs  the intuition that one must count the number of executions at bid when the system was in the state $(q^b,s)$ and normalize this quantity by the time spent in the state $(q^b,s)$. This is justified mathematically as follows. For any $(q^b,s=i\delta)$ $\in$ $\Qc^b\times\S$, let us define the point process
\beqs
N_t^{b,q^b,i} &=& \int_0^t 1_{\lbrace Q_u^b = q, S_{u-}= i\delta  \rbrace} dN_u^b, \;\;\; t \geq 0,
\enqs
which counts the number of jumps of $N^b$ when $(Q^b,S)$ was in state $(q^b,s=i\delta)$. 
Then, for any nonnegative predictable process $(H_t)$, we have
\beq
\Eb\Big[ \int_0^\infty H_t dN_t^{b,q^b,i}\Big] &=&  \Eb\Big[ \int_0^\infty H_t  1_{\lbrace Q_t^b = q^b, S_{t-}= i\delta  \rbrace}  dN_t^{b}\Big]  \nonumber \\
&=&  \Eb\Big[ \int_0^\infty H_t  1_{\lbrace Q_t^b = q^b, S_{t-}= i\delta  \rbrace}   \lambda^b(Q_t^b,S_t) dt  \Big]  \nonumber \\
&=&  \Eb\Big[ \int_0^\infty H_t  1_{\lbrace Q_t^b = q^b, S_{t-}= i\delta  \rbrace}   \lambda_i^b(q^b) dt  \Big],   \label{intensNb}
\enq
where we used in the second equality the fact that $\lambda^b(Q_t^b,S_t)$ is the intensity of $N^b$.   The relation \reff{intensNb} means  that the point process $N^{b,q^b,i}$ admits for intensity 
$\lambda_i^b(q^b)1_{\lbrace Q_t^b = q^b, S_{t-}= i\delta  \rbrace}$.  By defining
\beqs
\Tc^{b,q^b,i}_t &=& \int_0^t 1_{\lbrace Q_u^b = q, S_{u-}= i\delta  \rbrace} du 
\enqs
as  the time that $(Q^b,S)$ spent in the state $(q^b,s=i\delta)$, this means equivalently 
that  the process $M_t^{b,q^b,i}$ $=$ $N^{b,q^b,i}_{A_t^{b,q^b,i}}$, where $A_t^{b,q^b,i}$ $=$ $\inf\{ u \geq 0: \Tc_u^{b,q^b,i} \geq t\}$ is the 
c\`ad-l\`ag inverse of $\Tc^{b,q^b,i}$, is a Poisson process with intensity $\lambda_i(q^b)$.  By  assuming that  $\Tc^{b,q^b,i}_T$ is large  w.r.t. $\lambda_i(q^b)$, which is the case when $(\hat S_n)$ is irreducible (hence recurrent), and for high-frequency data over $[0,T]$, we have a consistent estimator of  $\lambda_i^b(q^b)$ given by: 
\beq \label{consistentestimatorbid}
\hat \lambda_i^b(q^b) &=&  \frac{N_T^{b,q^b,i}}{\Tc_T^{b,q^b,i}}. 
\enq
Similarly, we have a consistent estimator for  $\lambda_i^a(q^a)$ given by: 
\beq \label{consistentestimatorask}
\hat \lambda_i^a(q^a) &=&  \frac{N_T^{a,q^a,i}}{\Tc_T^{a,q^a,i}},   
\enq
where $N_T^{a,q^a,i}$ counts the number of executions at ask quote $q^a$ and for a spread $i\delta$, and $\Tc_T^{a,q^a,i}$ is the time that $(Q^a,S)$ spent in the state $(q^a,s=i\delta)$ over  $[0,T]$.

Let us now illustrate this estimation procedure on real data, with the same market data as above, i.e.  tick-by-tick level 1 for SOGN.PA on April 18, 2011, provided by Quant\-house via OneTick timeseries database.  Actually, since we did not perform the strategy on this real-world order book, we could not observe the real execution processes $N^{b}$ and $N^a$.   We built thus 
simple proxies $\tilde{N}^{b,q^b,i}$ and $\tilde{N}^{a,q^a,i}$,  for $q^b$ $=$ $Bb,Bb_+$, $q^a$ $=$ $Ba,Ba_-$, $i$ $=$ $1,\ldots,m$,  based on the following rules. Let us also assume that in addition to 
$(S_{\theta_n})_{n}$, we observe at jump times $\theta_n$ of the spread,  the volumes $(V^a_{\theta_n},V^b_{\theta_n})$  offered at the best ask  and best bid price in the LOB  together with  the cumulated market order quantities $\vartheta_{\theta_{n+1}}^{BUY}$  and 
$\vartheta_{\theta_{n+1}}^{SELL}$ 
arriving between two consecutive jump times $\theta_n$ and $\theta_{n+1}$ of the spread, respectively at best ask price and best bid price.    
We finally fix an arbitrarily  typical volume $V_0$, e.g. $V_0$ $=$ $100$ of our limit orders, and  define the proxys 
$\tilde{N}^{b,q^b,i}$ and $\tilde{N}^{a,q^a,i}$ at times $\theta_n$ by: 
\beqs
%\overline{\xi}_{\theta_{n+1}}^{SELL} &:=& \sum_{\theta_{n} \leq \vartheta_k^{SELL} < \theta_{n+1} } \xi_k^{SELL}\\
%\overline{\xi}_{\theta_{n+1}}^{BUY} &:=& \sum_{\theta_{n} \leq \vartheta_k^{BUY} < \theta_{n+1} } \xi_k^{BUY}\\
\tilde{N}^{b,Bb_+,i}_{\theta_{n+1}} &=& \tilde{N}^{b,Bb_+,i}_{\theta_{n}} 
+ 1_{\left\lbrace V_0 < \vartheta_{\theta_{n+1}}^{SELL}, S_{\theta_n}=i\delta  \right\rbrace} \,\, ,\, \tilde{N}^{b,Bb_+,i}_{0}=0 \\
\tilde{N}^{b,Bb,i}_{\theta_{n+1}} &=& \tilde{N}^{b,Bb_+,i}_{\theta_{n}} + 1_{\left\lbrace V_0+V^b_{\theta_n} < \vartheta_{\theta_{n+1}}^{SELL}, 
S_{\theta_n}=i\delta \right\rbrace}\,\, ,\,\tilde{N}^{b,Bb,i}_0=0\\
\tilde{N}^{a,Aa_-,i}_{\theta_{n+1}} &=& \tilde{N}^{a,Aa_-,i}_{\theta_{n}} + 1_{\left\lbrace V_0 < \vartheta_{\theta_{n+1}}^{BUY}, 
S_{\theta_n}=i\delta  \right\rbrace}\,\, ,\,\tilde{N}^{a,Aa_-,i}_0=0\\
\tilde{N}^{a,Aa,i}_{\theta_{n+1}} &=& \tilde{N}^{b,Aa,i}_{\theta_{n}} + 1_{\left\lbrace V_0+V^a_{\theta_n} < \vartheta_{\theta_{n+1}}^{BUY}, 
S_{\theta_n}=i\delta \right\rbrace}\,\, ,\, \tilde{N}^{a,Aa,i}_0=0, 
\enqs
together with a proxy for the time spent in spread $i\delta$: 
\beqs
\tilde \Tc_{\theta_{n+1}}^i &=&  \tilde \Tc_{\theta_{n}}^i  + (\theta_{n+1}-\theta_n) 1_{\{S_{\theta_n}=i\delta\}}, \;\;\; \tilde\Tc_0^i = 0.
\enqs
The interpretation of these proxies is the following: we consider the case where the (small) market maker instantaneously updates her quote $Q^b$ (resp. $Q^a$) and volume $L^b \leq V_0$ (resp. $L^a \leq V_0$) only when the spread changes exogenously, i.e. at dates $(\theta_n)$, so that the spread remains constant between her updates, not considering her own quotes. If she chooses to improve best price i.e $Q_{\theta_n}^b=Bb_+$ (resp. $Q_{\theta_n}^a=Ba_-$) she will be in top priority in the LOB and therefore captures all incoming market order flow to sell (resp. buy). Therefore, an unfavourable way  for (under)-estimating  her number of executions is to increment $\tilde{N}^b$ (resp. $\tilde{N}^a$) only when total traded volume at bid $\xi_{\theta_{n+1}}^{SELL}$ (resp. total volume traded at ask $\xi_{\theta_{n+1}}^{BUY}$) was greater than $V_0$. If the market maker chooses to add liquidity to the best prices i.e. $Q_{\theta_n}^b=Bb$ (resp. $Q_{\theta_n}^a=Ba$), she will be ranked behind $V^b_{\theta_n}$ (resp. $V^a_{\theta_n}$) in LOB priority queue. Therefore, we increment $\tilde{N}^b$ (resp. $\tilde{N}^a$) only when the total traded volume at bid 
$\vartheta_{\theta_{n+1}}^{SELL}$ (resp. total volume traded at ask $\vartheta_{\theta_{n+1}}^{BUY}$) was greater than $V_0+V^b_{\theta_n}$ (resp. $V_0+V^a_{\theta_n}$). We then  provide a proxy estimate  for $\lambda_i^b(q^b)$, $\lambda_i^a(q^a)$ by:
\beq \label{estimproxy}
\tilde \lambda_i^b(q^b) \; = \; \frac{\tilde N^{b,q^b,i}_{\theta_n}}{\tilde \Tc_{\theta_n}^i}, & & 
\tilde \lambda_i^a(q^a) \; = \; \frac{\tilde N^{a,q^a,i}_{\theta_n}}{\tilde \Tc_{\theta_n}^i}.
\enq

We performed the estimation procedure \reff{estimproxy}.  In table \ref{execintensities} we computed $\tilde{\lambda}_i^a(q^a)$ and 
$\tilde{\lambda}_i^b(q^b)$ for $i$ $=$ $1,\ldots,6$, and limit order quotes $q^b$ $=$ $Bb_+,Bb$, $q^a$ $=$ $Ba,Ba_-$.  Due to the lack of data, estimate for large values of the spread are less robust. 
In this table, each row corresponds to a choice of  the spread  and each column to a choice of  the quotes. 

\begin{table}[h!]
\footnotesize
\begin{center}
\begin{tabular}{|l	|	l		l		l		l	|}

\hline	&	Ba	&	Ba-	&	Bb	&	Bb+	\\
\hline 0.005	&	0.0539	&	0.1485	&	0.0718	&	0.1763	\\
0.010	&	0.0465	&	0.0979	&	0.0520	&	0.1144	\\
0.015	&	0.0401	&	0.0846	&	0.0419	&	0.0915	\\
0.020	&	0.0360	&	0.0856	&	0.0409	&	0.0896	\\
0.025	&	0.0435	&	0.1009	&	0.0452	&	0.0930	\\
0.030	&	0.0554	&	0.1202	&	0.0614	&	0.1255	\\
\hline 
\end{tabular}
\end{center}
\caption{Estimate of execution intensities $\tilde{\lambda}_i^a(q^a)$ and $\tilde{\lambda}_i^b(q^b)$, expressed in $s^{-1}$ for 
SOGN.PA on April 18, 2011,  between 9:30 and 16:30 Paris local time. Each row corresponds to a spread value as a multiple of $\delta=0.005$. Each column corresponds  to a limit order quote.}
\label{execintensities}
\end{table}

\begin{figure}[h!] 
\centering
\includegraphics[width=0.9\textwidth]{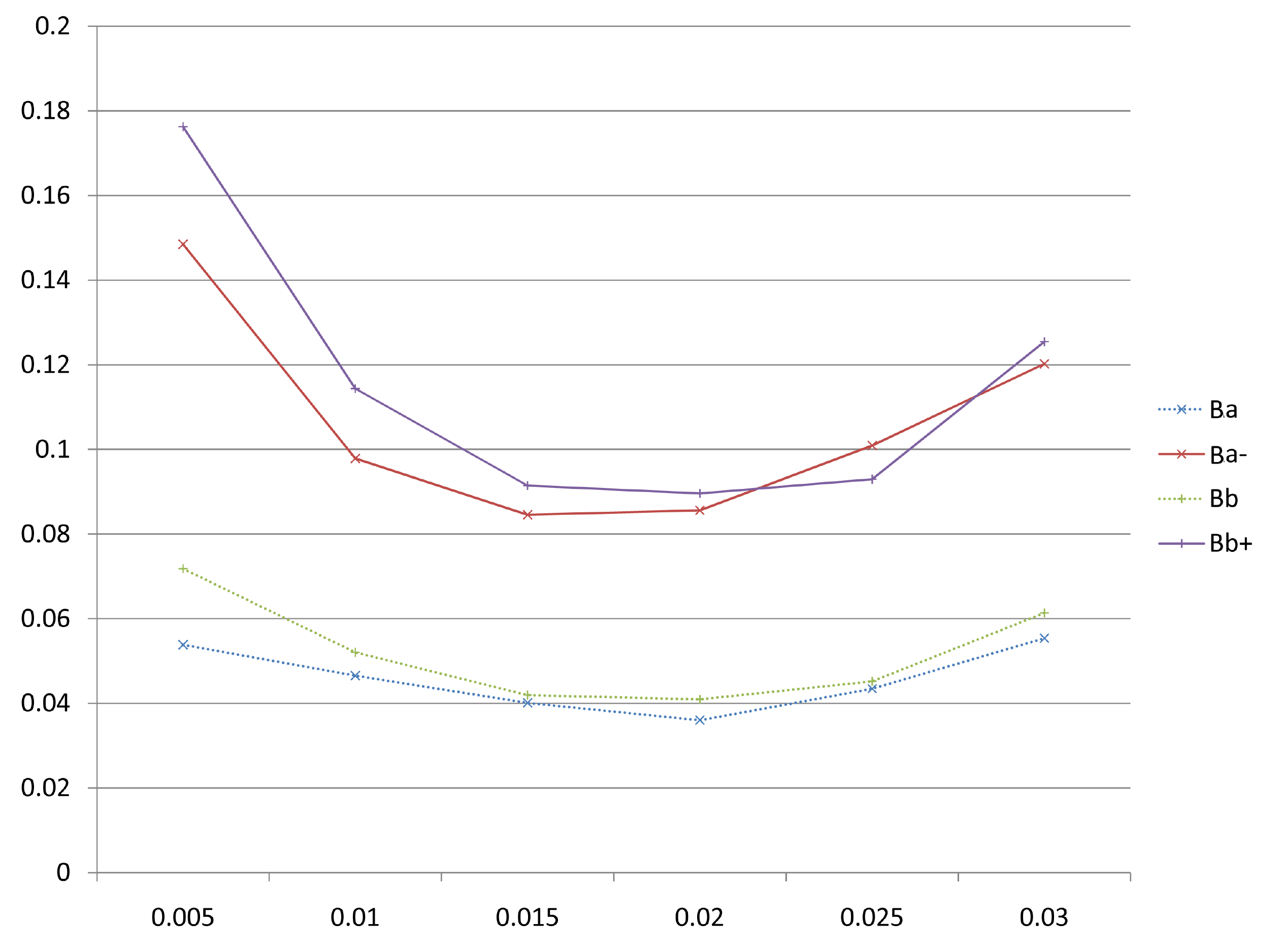}
\caption{Plot of execution intensities estimate as a function of the spread for the stock SOGN.PA on the 18/04/2011, expressed in $s^{-1}$ (affine interpolation).}
\label{execintensitiesplot}
\end{figure}

In figure \ref{execintensitiesplot} we plotted this estimated intensity as a function of the spread, i.e. $s$ $=$ $i\delta$ 
$\rightarrow$ $\tilde\lambda^b_i(q^b)$, $\tilde\lambda^a_i(q^a)$ for $q^b$ $\in$ $\Qc^b$, and $q^a$ $\in$ $\Qc^a$. As one would expect,  $(\tilde\lambda^a_i(.),\tilde\lambda^b_i(.))$ are decreasing functions of $i$ for the small values of $i$ which matches the intuition that the higher are the (indirect) costs, the smaller is market order flow. Surprisingly, for large values of $i$ this function becomes increasing, which can be due either to an estimation error, caused by the lack of data for this spread range, or a ``gaming" effect, in other word liquidity providers increasing their spread when large or autocorrelated market orders come in.

% at date $\theta_n$ the spread process jumps, and thus we update the limit prices $Q^b_{\theta_n}$ (resp. $Q^a_{\theta_n}$). If $Q^b_{\theta_n}=Bb_+$ (resp. $Q^a_{\theta_n}=Aa_-$) this order has the highest priority in the order book, therefore, it will be fully executed before the spread moves (at date $\theta_{n+1}$) iff the total traded volume on the bid side (resp. ask side) is greater than $L^b_{\theta_n}$ (resp. $L^a_{\theta_n}$). Otherwise, if $Q^b_{\theta_n}=Bb$ (resp. $Q^a_{\theta_n}=Aa$) this order's priority is lower than the existing volume at best bid (resp. best ask). Therefore, it will be fully executed before the spread moves (at date $\theta_{n+1}$) iff the total traded volume on the bid side (resp. ask side) is greater than $L^b_{\theta_n}$ plus the existing volume at best bid at date $\theta_n$ (resp. $L^a_{\theta_n}$ plus the existing volume at best bid at date $\theta_n$). In all our following computations, we supposed that $L^b = L^a =100$ to simplify computations. More precisely we computed the following quantities:
%\beqs
%\hat{N}^{b,Bb_+,s}_{\theta_{n+1}}= \hat{N}^{b,Bb_+,s}_{\theta_{n}}
%\enqs
 
\section{Optimal limit/market order  strategies}

\setcounter{equation}{0} \setcounter{Assumption}{0}
\setcounter{Theorem}{0} \setcounter{Proposition}{0}
\setcounter{Corollary}{0} \setcounter{Lemma}{0}
\setcounter{Definition}{0} \setcounter{Remark}{0}

\subsection{Control problem formulation}

Our  market model in the previous section is completely determined by the state variables $(X,Y,P,S)$ controlled by the limit/marker order trading strategies $\alpha$ $\in$ $\Ac$. 

The objective of the market maker is the following. She wants to maximize over a finite horizon $T$ the profit from her transactions in the LOB, while 
keeping under control her inventory (usually starting from zero), and getting rid of her inventory at the terminal date: 
\beq \label{criterion}
\mbox{ maximize } \;\;\;  \Eb\big[ U(X_T) - \gamma \int_0^T g(Y_t) dt \big]
\enq  
over all limit/market order trading strategies $\alpha$ $=$ $(\alpha^{make},\alpha^{take})$ in $\Ac$ such that $Y_T$ $=$ $0$. Here $U$ is an increasing reward function, $\gamma$ is a nonnegative constant, and $g$ is a nonnegative convex function, so that  the last integral term $\int_0^T g(Y_t)dt$ penalizes the variations of the inventory.  Typical frameworks  include the two following cases:

\vspace{1mm}

$\bullet$ Mean-quadratic criterion: $U(x)$ $=$ $x$, $\gamma$ $>$ $0$, $g(y)$ $=$ $y^2$. 

\vspace{1mm}

$\bullet$ Exponential utility maximization: $U(x)$ $=$  $-\exp(-\eta x)$, $\gamma$ $=$ $0$. 

\vspace{1mm}

Let us first remove mathematically the terminal constraint  on the inventory: $Y_T$ $=$ $0$, by introducing the liquidation function 
$L(x,y,p,s)$ defined on $\R^2\times\P\times\S$ by:
\beqs
L(x,y,p,s) &=& x - c(-y,p,s) \; = \;   x + yp - |y|\frac{s}{2}- \eps. 
\enqs 
This represents the value that an investor would obtained by liquidating immediately by a market order her  inventory position $y$ in stock, given a cash holdings $x$, a mid-price $p$ and a spread $s$. Then, problem \reff{criterion} is formulated equivalently as
\beq \label{criterion2}
\mbox{ maximize } \;\;\;  \Eb\big[ U(L(X_T,Y_T,P_T,S_T)) - \gamma \int_0^T g(Y_t) dt \big]
\enq  
over all limit/market order trading strategies $\alpha$ $=$ $(\alpha^{make},\alpha^{take})$ in $\Ac$. Indeed, the maximal value of problem 
\reff{criterion} is clearly smaller than the one of problem \reff{criterion2} since for any $\alpha$ $\in$ $\Ac$ s.t. $Y_T$ $=$ $0$, we have 
$L(X_T,Y_T,P_T,S_T)$ $=$ $X_T$. Conversely, given an arbitrary $\alpha$ $\in$ $\Ac$, let us consider the control $\tilde\alpha$ $\in$ $\Ac$,  coinciding with $\alpha$ up to time $T$, 
and to which one add at the  terminal date $T$  the market order consisting in liquidating all the inventory $Y_T$. 
The associated state process $(\tilde X,\tilde Y,P,S)$ satisfies: $\tilde X_t$ $=$ $X_t$, $\tilde Y_t$ $=$ $Y_t$ for $t$ $<$ $T$, and 
$\tilde X_T$ $=$ $L(X_T,Y_T,P_T,S_T)$, $\tilde Y_T$ $=$ $0$. This shows that the maximal value of problem \reff{criterion2} is smaller and then equal to the maximal value of problem \reff{criterion}. 

We then define the value function for problem \reff{criterion2} (or \reff{criterion}): 
\beq \label{defvalue}
v(t,z,s) &=& \sup_{\alpha\in\Ac}  \Eb_{t,z,s}  \Big[ U(L(Z_T,S_T)) - \gamma \int_t^T g(Y_u) du  \Big]
\enq
for  $t$ $\in$ $[0,T]$,  $z$ $=$ $(x,y,p)$ $\in$ $\R^2\times\P$, $s$ $\in$ $\S$. Here, given $\alpha$ $\in$ $\Ac$, $\Eb_{t,z,s}$ denotes the expectation operator under which the process $(Z,S)$ $=$ $(X,Y,P,S)$ solution to  \reff{dynS}-\reff{dynY}-\reff{dynX}-\reff{sautY}-\reff{sautX}, 
with initial state $(Z_{t^-},S_{t^-})$ $=$ $(z,s)$, is taken.  Problem \reff{defvalue} is a mixed regular/impulse control problem in a 
regime switching  jump-diffusion model, that we shall study by dynamic programming methods.  Since the spread takes finite values in 
$\S$ $=$ $\delta\I_m$, it will be convenient to denote for  $i$ $\in$ $\I_m$, by $v_i(t,z)$ $=$ $v(t,z,i\delta)$. By misuse of notation, we shall often 
identify the value function  with the $\R^m$-valued function $v$ $=$ $(v_i)_{i\in\I_m}$ defined on $[0,T]\times\R^2\times\P$.

\subsection{Dynamic programming equation}

For any $q$ $=$ $(q^b,q^a)$ $\in$ $\Qc$, $\ell$ $=$ $(\ell^b,\ell^a)$ $\in$ $[0,\bar\ell]^2$,  we consider the second-order nonlocal operator: 
\beq
\Lc^{q,\ell} \varphi(t,x,y,p,s) &=& \Pc \varphi(t,x,y,p,s)  +  R(t) \varphi(t,x,y,p,s) \nonumber \\
& &  \; + \;  \lambda^b(q^b,s) \big[  \varphi(t,\Gamma^b(x,y,p,s,q^b,\ell^b),p,s) - \varphi(t,x,y,p,s) \big] \nonumber \\
& &  \; + \;  \lambda^a(q^a,s) \big[  \varphi(t,\Gamma^a(x,y,p,s,q^a,\ell^a) ,p,s) - \varphi(t,x,y,p,s) \big],  \label{defLc}
\enq
for $(t,x,y,p,s)$ $\in$ $[0,T]\times\R^2\times\P\times\S$, where 
\beqs
R(t) \varphi(t,x,y,p,s) &=& \sum_{j=1}^m r_{ij}(t)\big[ \varphi(t,x,y,p,j\delta) - \varphi(t,x,y,p,i\delta) \big], \;\;\; \mbox{ for } s = i \delta, \; i \in \I_m, 
\enqs
and $\Gamma^b$ (resp. $\Gamma^a$) is defined from $\R^2\times\P\times\S\times\Qc^b\times\R_+$ (resp. 
$\R^2\times\P\times\S\times\Qc^a\times\R_+$ into $\R^2$) by
\beqs
\Gamma^b(x,y,p,s,q^b,\ell^b) &=& (x-\pi^b(q^b,p,s) \ell^b,y+\ell^b) \\
\Gamma^a(x,y,p,s,q^a,\ell^a) &=& (x+ \pi^a(q^a,p,s) \ell^a,y- \ell^a). 
\enqs 
The  first term of $\Lc^{q,\ell}$ in \reff{defLc} correspond  to the infinitesimal generator of the diffusion mid-price process $P$, the second one is the generator of the continuous-time spread Markov chain  $S$, and the two last terms correspond to the nonlocal operator induced by the jumps of the cash process $X$ and inventory process $Y$ when applying an instantaneous limit order control $(Q_t,L_t)$ $=$ $(q,\ell)$.  
 
Let us also consider the impulse operator associated to market order control, and defined by 
\beqs
\Mc \varphi(t,x,y,p,s) &=& \sup_{e \in [-\bar e,\bar e]} \varphi(t,\Gamma^{take}(x,y,p,s,e),p,s),
\enqs
where  $\Gamma^{take}$ is the impulse transaction function  defined from 
$\R^2\times\P\times\S\times\R$ into $\R^2$ by:  
\beqs
\Gamma^{take}(x,y,p,s,e) &=& \Big(x - c(e,p,s), y + e\Big),
\enqs

The dynamic programming equation (DPE) associated to the control problem \reff{defvalue} is the quasi-variational inequality (QVI): 
\beq \label{dynv}
\min\big[ - \Dt{v} - \sup_{(q,\ell)\in\Qc(s)\times[0,\bar\ell]^2} \Lc^{q,\ell} v \; + \;  \gamma g  \; , \; v - \Mc v \big] &=& 0,   \mbox{ on }  
[0,T)\times\R^2\times\P\times\S
\enq
together with the terminal condition:
\beq \label{termv}
v(T,x,y,p,s) &=& U(L(x,y,p,s)), \; \forall (x,y,p) \in  \R^2\times\P\times\S. 
\enq  
This is  also written explicitly in terms of system of QVIs for  the functions $v_i$, $i$ $\in$ $\I_m$:
\beqs
\min\Big[ - \Dt{v_i}   -  \Pc v_i -    \sum_{j=1}^m r_{ij}(t) [ v_j(t,x,y,p) - v_i(t,x,y,p)] & & \\
\;\;\;\;\; -\; \sup_{(q^b,\ell^b)\in\Qc_i^b\times[0,\bar\ell]} \lambda_i^b(q^b) [v_i(t,x-\pi_i^b(q^b,p)\ell^b,y+\ell^b,p) -v_i(t,x,y,p)] & & \\
\;\;\;\;\; -\; \sup_{(q^a,\ell^a)\in\Qc_i^a\times[0,\bar\ell]} \lambda_i^a(q^a) [v_i(t,x+\pi_i^a(q^a,p)\ell^a,y- \ell^a,p) -v_i(t,x,y,p)]  \; + \gamma g(y) \; ;   & & \\
 v_i(t,x,y,p)   - \sup_{e\in [-\bar e,\bar e]} v_i(t,x-c_i(e,p),y+e,p) \Big] & =& 0, 
\enqs 
for $(t,x,y,p)$ $\in$  $[0,T)\times\R^2\times\P$, together with the terminal condition:
\beqs
v_i(T,x,y,p) &=& U(L_i(x,y,p)), \;\;\;  \;\;\; \forall (x,y,p) \in  \R^2\times\P,
\enqs 
where we set $L_i(x,y,p)$ $=$ $L(x,y,p,i\delta)$.

By methods of dynamic programming,  one can show by standard arguments that the value function $v$ is the  unique viscosity solution to the QVI 
\reff{dynv}-\reff{termv} under suitable growth conditions depending on the utility function $U$ and penalty function $g$. 
We next focus on some particular cases of interest  for reducing remarkably  the number of states variables in the dynamic programming equation DPE. 

% A VERIFIER RIGOUREUSEMENT. 

%The next section is devoted to numerical schemes for the resolution of the dynamic programming equation  DPE \reff{dynv}-\reff{termv}, and to %some particular cases of interest for reducing remarkably  the number of states variables in the DPE. 
 
\subsection{Two special cases}

{\bf 1. Mean criterion with penalty on inventory} 
 
\vspace{1mm}

\noindent  We first  consider the case as in \cite{stosag09}  where:
\beq \label{casemar}
U(x)   \; = \;  x, \;\; x \in \R, & \mbox{ and } \;\;\;  (P_t)_t  \; \mbox{ is a martingale}. 
\enq
The martingale assumption of the stock price under the historical measure under which the market maker performs her criterion, reflects the idea that she has no information on the future direction of the stock price.  Moreover, by starting typically from zero endowment in stock, and by introducing a penalty function on inventory,  the market maker  wants to keep an inventory that fluctuates around zero.    

In this case, and exploiting similarly as in \cite{baylud11}  the martingale property that $\Pc p$ $=$ $0$, we see that  the solution 
$v$ $=$ $(v_i)_{i\in\I_m}$ to the dynamic programming system  \reff{dynv}-\reff{termv} is reduced into the form: 
\beq \label{simpli}
v_i(t,x,y,p) &=& x + yp + \phi_i(t,y),
\enq
where $\phi$ $=$ $(\phi_i)_{i\in\I_m}$ is solution the system of  integro-differential equations (IDE): 
 \beqs
\min\Big[ - \Dt{\phi_i}    -    \sum_{j=1}^m r_{ij}(t) [ \phi_j(t,y) - \phi_i(t,y)] & & \\
\;\;\;\;\; -\; \sup_{(q^b,\ell^b)\in\Qc_i^b\times[0,\bar\ell]} \lambda_i^b(q^b) [\phi_i(t,y+\ell^b) -\phi_i(t,y)  
+ \big( \frac{i\delta}{2} - \delta 1_{q^b=Bb_+} \big)\ell^b ] & & \\
\;\;\;\;\; -\; \sup_{(q^a,\ell^a)\in\Qc_i^a\times[0,\bar\ell]} \lambda_i^a(q^a)  [\phi_i(t,y- \ell^a) -\phi_i(t,y)  
+ \big( \frac{i\delta}{2} - \delta 1_{q^a=Ba_-} \big)\ell^a ]  \; + \gamma g(y) \; ;   & & \\
 \phi_i(t,y)   - \sup_{e\in [-\bar e,\bar e]} [ \phi_i(t,y+e) - \frac{i\delta}{2}|e| - \eps  ] \Big] & =& 0, 
\enqs 
together with the terminal condition: 
\beqs
\phi_i(T,y) &=& -|y|\frac{i\delta}{2} - \eps,
\enqs
These one-dimensional IDEs 
can be solved numerically  by a standard finite-difference scheme by discretizing the time derivative of $\phi$, and the grid 
space in $y$.  The reduced form \reff{simpli} shows that the optimal market making strategies are price independent, and depend only on the level of inventory and of the spread, which is consistent with stylized features in the market.  Actually,  the IDEs for  $(\phi_i)$ even show  
that optimal policies do not depend on the martingale modeling of the stock price.

\vspace{2mm}

\noindent  {\bf 2.  Exponential utility criterion}

\vspace{1mm}

\noindent We  next  consider as in \cite{avesto08} a risk averse market marker:  
\beq 
U(x)  &=&  -\exp(-\eta x),  \; x \in \R, \; \eta > 0, \;\;\; \gamma \; = \;  0, 
\enq
and assume that $P$ follows a Bachelier model:
\beqs
dP_t &=& b dt +  \sigma dW_t.
\enqs
Such  price process may take  negative values in theory, but at the short-time horizon where high-frequency trading take place,  the evolution 
of an arithmetic Brownian motion looks very similar to a geometric Brownian motion as in the Black-Scholes model.  

In this case,  we see, similarly as in \cite{gueferleh11},  that  the solution $v$ $=$ $(v_i)_{i\in\I_m}$ to the dynamic programming system  
\reff{dynv}-\reff{termv} is reduced into the form: 
\beq \label{simpliexp}
v_i(t,x,y,p) &=& U(x + yp) \varphi_i(t,y),
\enq
where $\varphi$ $=$ $(\varphi_i)_{i\in\I_m}$ is solution the system of one-dimensional integro-differential equations (IDE): 
 \beqs
\max\Big[ - \Dt{\varphi_i}   + (b\eta y - \frac{1}{2} \sigma^2(\eta y)^2)\varphi_i -    \sum_{j=1}^m r_{ij}(t) [ \varphi_j(t,y) - \varphi_i(t,y)] & & \\
\;\;\;\;\; -\; \inf_{(q^b,\ell^b)\in\Qc_i^b\times[0,\bar\ell]} \lambda_i^b(q^b) [  \exp\big(-\eta \big(\frac{i\delta}{2}  - \delta 1_{q^b=Bb_+}\big)\ell^b\big) 
\varphi_i(t,y+\ell^b) -\varphi_i(t,y) ] & & \\
\;\;\;\;\; -\; \inf_{(q^a,\ell^a)\in\Qc_i^a\times[0,\bar\ell]} \lambda_i^a(q^a)  [  \exp\big(-\eta \big(\frac{i\delta}{2} -   \delta 1_{q^a=Ba_-}\big)\ell^a \big)
\varphi_i(t,y- \ell^a) -\varphi_i(t,y)  ]    & & \\
 \varphi_i(t,y)   - \inf_{e\in [-\bar e,\bar e]} [  \exp\big(\eta |e|\frac{i\delta}{2} + \eta\eps \big) \varphi_i(t,y+e) ] \Big] & =& 0, 
\enqs 
together with the terminal condition: 
\beqs
\varphi_i(T,y) &=& \exp(\eta |y| \frac{i \delta}{2}). 
\enqs
Actually, we notice that the reduced form \reff{simpliexp}  holds more generally when $P$ is a L\'evy process, by  using the property that in the case: 
$\Pc U(x+yp)$ $=$ $\psi(y) U(x+yp)$ for some  function $\psi$ depending on the characteristics of the generator $\Pc$ of the L\'evy  process. 
As in the above mean-variance criterion case, the reduced form \reff{simpliexp} shows that the optimal market making strategies are price independent, and depend only on the level of inventory and of the spread.  However, it depends on the model (typically the volatility) for the stock price.

%\begin{Remark}
%{\rm Mettre l'EDP verifie par $\varphi(t,y,s)$ en remplacant $v(t,y,p,s)$ $=$ $U(x+yp)\varphi(t,y,s)$ dans le DPE.
%}
%\end{Remark}

\section{Computational results}
 
\setcounter{equation}{0} \setcounter{Assumption}{0}
\setcounter{Theorem}{0} \setcounter{Proposition}{0}
\setcounter{Corollary}{0} \setcounter{Lemma}{0}
\setcounter{Definition}{0} \setcounter{Remark}{0} 

In this section, we provide numerical results  in the case of a mean criterion with penalty on inventory, that we will denote within this section by 
$\alpha^{\star}$. We used parameters shown in table \reff{Parameters} together with transition probabilities $(\rho_{ij})_{1\leq i,j\leq M}$ calibrated in table \reff{transitionmatrix} and execution intensities calibrated in table \reff{execintensities}, slightly modified to make the bid and ask sides symmetric. 

\begin{table}[h!]
\footnotesize
\begin{center}
\subfloat[Market parameters]{
\begin{tabular}{|l|l|l|}
\hline 
Parameter & Signification &Value \\ 
\hline 
$ \delta $ & Tick size &0.005 \\ 
$ \rho $ & Per share rebate &0.0008 \\ 
$ \epsilon $ & Per share fee &0.0012 \\ 
$ \epsilon_0 $ & Fixed fee & $10^{-6}$ \\ 
$ \lambda(t) $ & Tick time intensity & $\equiv 1 s^{-1}$ \\ 
\hline 
\end{tabular}
} 
\subfloat[Optimization parameters]{\begin{tabular}{|l|l|l|}
\hline 
Parameter & Signification &Value \\ 
\hline 
$U(x)$ & Utility function & $x$ \\ 
$ \gamma $ & Inventory penalization &5 \\ 
$ \bar\ell $ & Max. volume make &100 \\ 
$ \bar e $ & Max. volume take &100 \\ 
\hline 
\end{tabular}
} \\ 
\subfloat[Discretization/localization parameters]{\begin{tabular}{|l|l|l|}
\hline 
Parameter & Signification &Value \\ 
\hline 
$ T $ & Length in seconds &300 s \\ 
$ y_{min} $ & Lower bound shares &-1000 \\ 
$ y_{max} $ & Upper bound shares &1000 \\ 
$ n $ & Number of time steps &100 \\ 
$ m $ & Number of spreads &6 \\ 
\hline 
\end{tabular}
} 
\subfloat[Backtest parameters]{\begin{tabular}{|l|l|l|}
\hline 
Parameter & Signification &Value \\ 
\hline 
$ N^{MC} $ & Number of paths for MC simul. & $10^5$ \\ 
$ \Delta t $ & Euler scheme time step & 0.3 s\\
$ \bar\ell_0 $ & B/A qty for bench. strat. &100 \\ 
$ x_0 $ & Initial cash &0 \\ 
$ y_0 $ & Initial shares &0 \\ 
$ p_0 $ & Initial price &45 \\ 
\hline 
\end{tabular}
} 
\end{center}
\caption{Parameters}
\label{Parameters}
\end{table}

\vspace{2mm}

\noindent  {\bf  \em{Shape of the optimal policy}.} The reduced form \reff{simpli} shows that the optimal policy $\alpha^\star$ does only depend on time $t$, inventory $y$ and spread level $s$. One can represent $\alpha^\star$ as a mapping $\alpha^\star : \R^+\times \R\times \S \rightarrow \Ac$ with $\alpha^\star = (\alpha^{\star,make} ,\alpha^{\star,take} )$ thus it divides the space $\R^+\times \R\times \S$ in two zones $\mathcal{M}$ and $\mathcal{T}$ so that $\alpha^\star_{\mid \mathcal{M}} = (\alpha^{\star,make} ,0 )$ and $\alpha^\star_{\mid \mathcal{T}} = (0 ,\alpha^{\star,take})$. Therefore we plot the optimal policy in one plane, distinguishing the two zones by a color scale. For the zone $\mathcal{M}$, due to the complex nature of the control, which is made of four scalars, we only represent  the prices regimes.

\begin{figure}[h!] 
\centering
\subfloat[near date $0$]{
\includegraphics[width=0.5\textwidth]{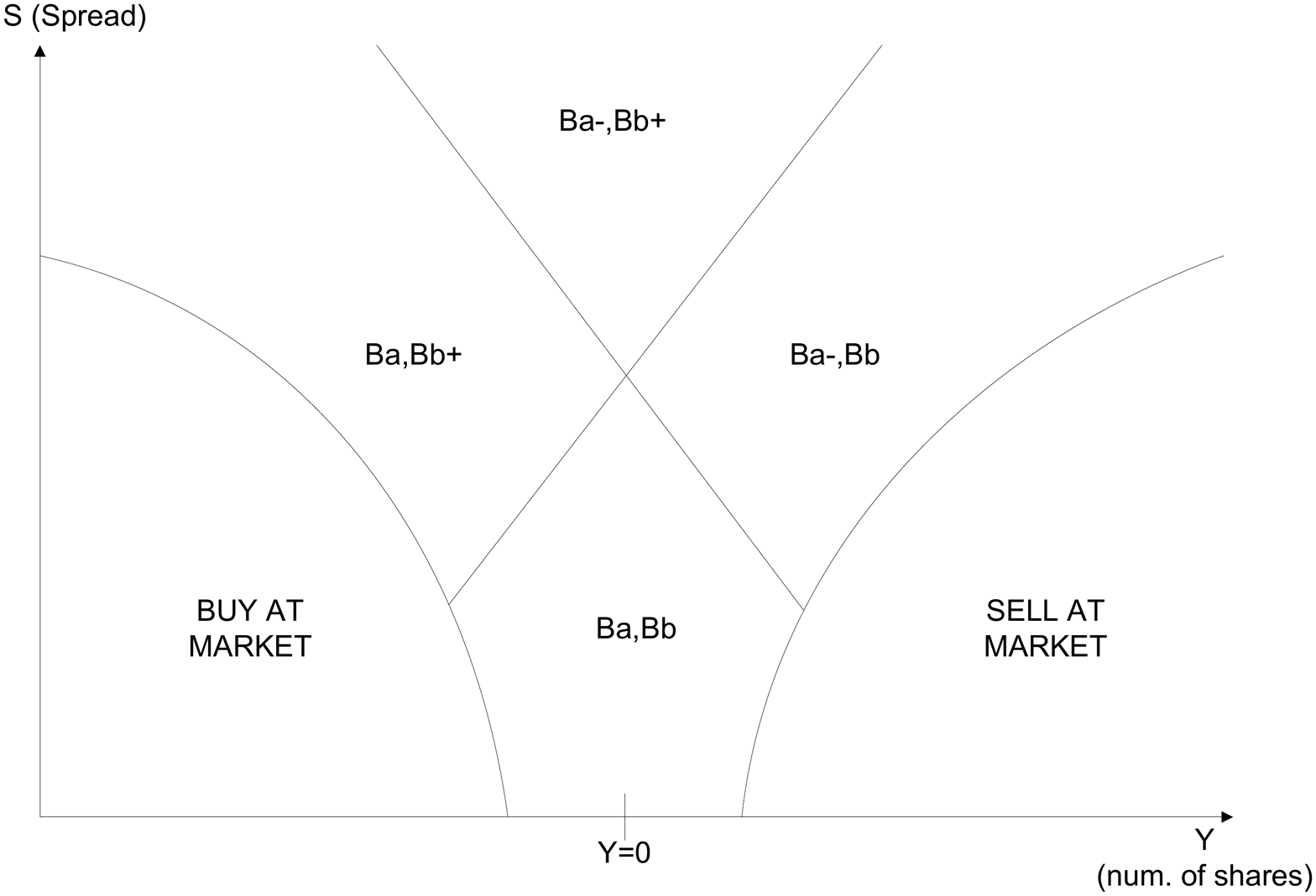}
}
\subfloat[near date $T$]{
\includegraphics[width=0.5\textwidth]{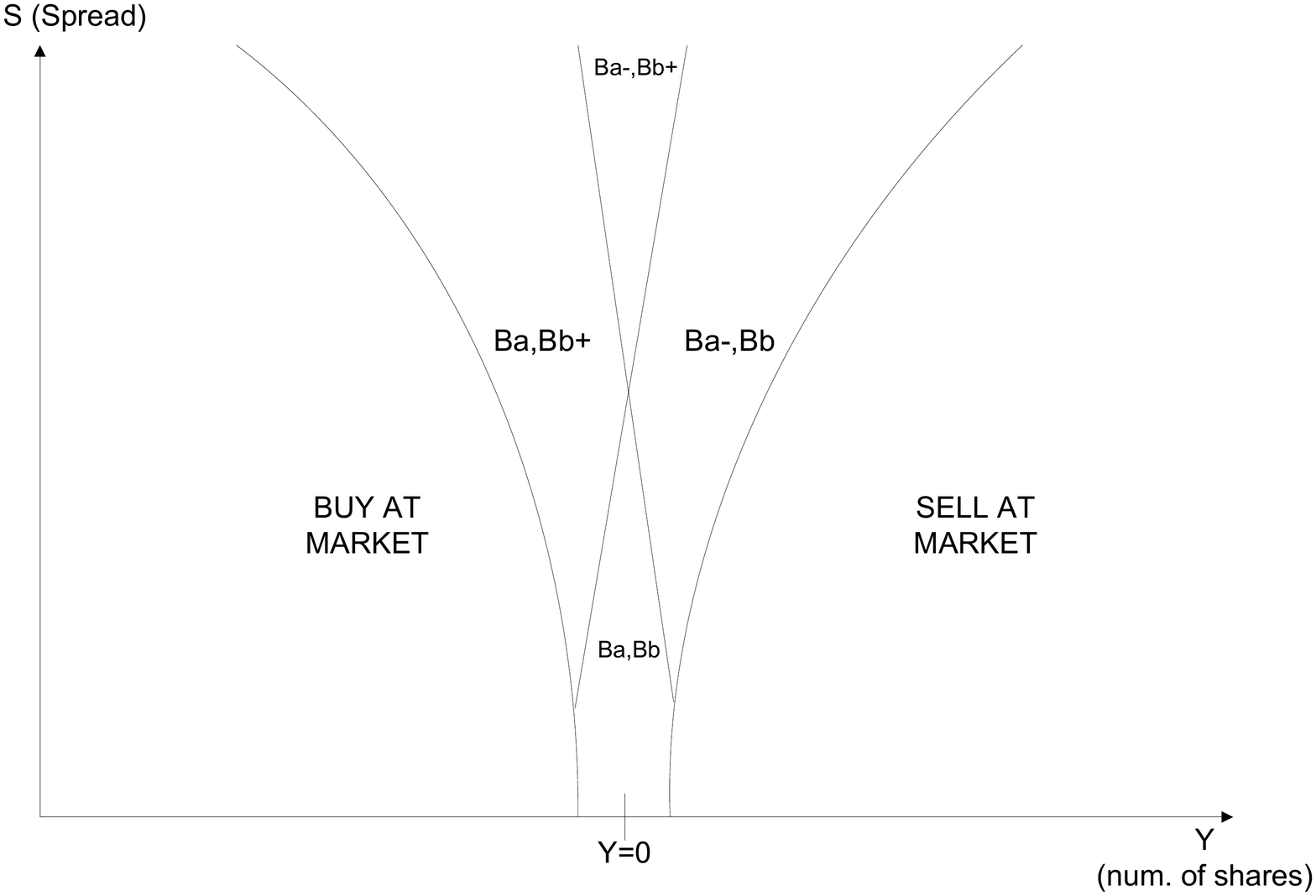}
}
\caption{Stylized shape of the optimal policy sliced in YS.}
\label{stylizedShapePolicyYS}
\end{figure}

Moreover, when using constant tick time intensity $\lambda(t) \equiv \lambda$ and in the case where $T\gg \frac{1}{\lambda}$ we can observe on numerical results that the optimal policy is mainly time invariant near date $0$; on the contrary, close to the terminal date $T$ the optimal policy has a transitory regime, in the sense that it critically depends on the time variable $t$. This matches the intuition that to ensure the terminal constraint $Y_T=0$, the optimal policy tends to get rid of the inventory more aggressively when close to maturity. In figure \ref{stylizedShapePolicyYS}, we plotted a stylized view of the optimal policy, in the plane $(y,s)$, to illustrate this phenomenon.

\vspace{2mm}

\noindent  {\bf  \em{Benchmarked empirical performance analysis}.} We made a backtest of the optimal strategy $\alpha^{\star}$, on simulated data, and benchmarked the results with the three following strategies:
\vspace{1mm}

$\bullet$ Optimal strategy without market orders (WoMO), that we denote by $\alpha^{w}$: this strategy is computed using the same IDEs,  but in the case where the investor is not allowed to use market orders, which is equivalent to setting 
$\bar e=0$.

\vspace{1mm}

$\bullet$ Constant strategy, that we denote by $\alpha^{c}$: this strategy is the symmetric best bid, best ask strategy with constant quantity $\bar\ell_0$ on both sides, or more precisely $\alpha^{c} := (\alpha^{c,make},0)$ with $\alpha_t^{c,make}\equiv (Bb,Ba,\bar\ell_0,\bar\ell_0)$.

\vspace{1mm}

$\bullet$ Random strategy, that we denote by $\alpha^{r}$: this strategy consists in choosing randomly the price of the limit orders and using constant quantities on both sides, or more precisely $\alpha^{r} := (\alpha^{r,make},0)$ with $\alpha_t^{r,make} = (\varsigma_t^b,\varsigma_t^a,\bar\ell_0,\bar\ell_0)$ where $(\varsigma_.^b,\varsigma_.^a)$ is s.t. $\forall t\in [0;T]$ ,  $\Pb(\varsigma_t^b=Bb)=\Pb(\varsigma_t^b=Bb_+)=\Pb(\varsigma_t^a=Ba)=\Pb(\varsigma_t^a=Ba_-)=\frac{1}{2}$.

\vspace{1mm}

Our backtest procedure is described as follows. For each strategy $\alpha \in \lbrace \alpha^{\star},\alpha^{w},\alpha^{c},\alpha^{r} \rbrace$, we simulated $N^{MC}$ paths of the tuple $(X^{\alpha},Y^{\alpha},P,S,N^{a,\alpha},N^{b,\alpha})$ on $[0,T]$, according to eq. \reff{dynS}-\reff{dynY}-\reff{dynX}-\reff{sautY}-\reff{sautX}, using a standard Euler scheme with time-step $\Delta t$. Therefore we  can  compute the empirical mean (resp. empirical standard deviation), that we denote by $m(.)$ (resp. $\sigma(.)$), for several quantities shown in table \reff{perfsynthesis}.

\begin{table}[h!]
\footnotesize
\begin{center}
\begin{tabular}{|l|l|l|l|l|l|}
\hline  & & optimal $\alpha^\star$ & WoMO $\alpha^w$& constant $\alpha^c$& random $\alpha^r$ \\ 
\hline 
Terminal wealth & $m(X_T)/\sigma(X_T)$ &2.117& 1.999& 0.472& 0.376\\ 
 & $m(X_T)$ &26.759& 25.19& 24.314& 24.022\\ 
 & $\sigma(X_T)$&12.634& 12.599& 51.482& 63.849\\ 
\hline Num. of exec. at bid & $m(N^b_T)$ & 18.770& 18.766& 13.758& 21.545\\ 
 & $\sigma(N^b_T)$& 3.660& 3.581& 3.682& 4.591\\ 
\hline Num. of exec. at ask & $m(N^a_T)$ & 18.770& 18.769& 13.76& 21.543\\ 
 & $\sigma(N^a_T)$& 3.666& 3.573& 3.692& 4.602\\ 
\hline Num. of exec. at market & $m(N^{market}_T)$& 6.336& 0& 0& 0\\ 
 &$\sigma(N^{market}_T)$ & 2.457& 0& 0& 0\\ 
\hline Maximum Inventory  & $m(\sup_{s\in [0;T]}\vert Y_s \vert)$ & 241.019& 176.204& 607.913& 772.361\\ 
 & $ \sigma(\sup_{s\in [0;T]}\vert Y_s \vert)$ & 53.452& 23.675 &272.631& 337.403\\ 
\hline 
\end{tabular}
\end{center}
\caption{Performance analysis: synthesis of benchmarked backtest ($10^5$ simulations).}
\label{perfsynthesis}
\end{table}

\begin{figure}[h!] 
\centering
\includegraphics[width=0.5\textwidth]{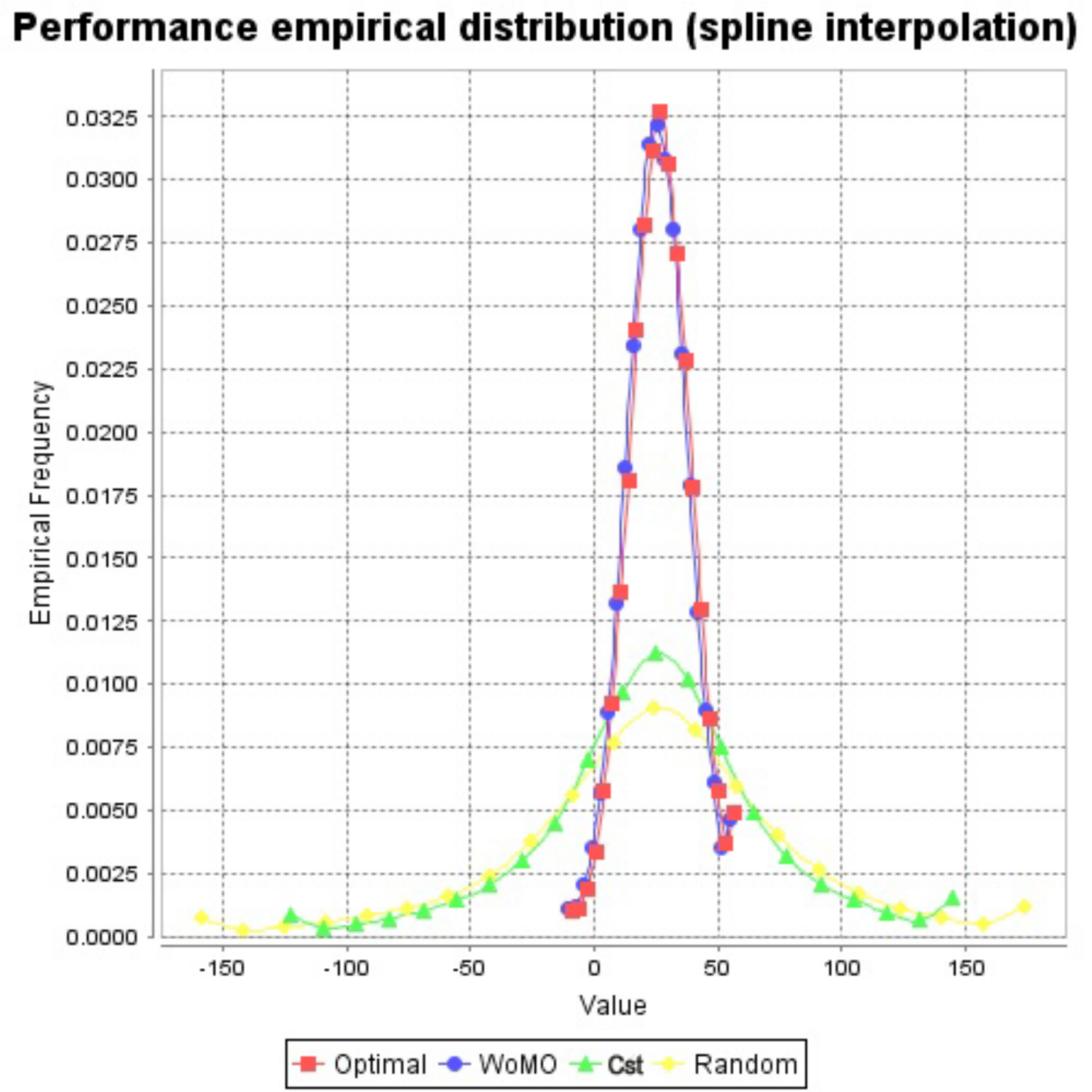} 
\caption{Empirical distribution of terminal wealth $X_T$ (spline interpolation).}
\label{empdistrib}
\end{figure}

Optimal strategy $\alpha^\star$ demonstrates significant improvement of the information ratio $\operatorname{IR}(X_T):= m(X_T)/\sigma(X_T)$ compared to the benchmark, which is confirmed by the plot of the whole empirical distribution of $X_T$ (see figure \reff{empdistrib}). 

Even if absolute values of $m(X_T)$ are not representative of what would be the real-world performance of such strategies, these results prove that the different layers of optimization are relevant. Indeed, one can compute the ratios $\left[m(X_T^{\alpha^\star})-m(X_T^{\alpha^c})\right]/\sigma(X_T^{\alpha^\star})= 0.194$ and $\left[ m(X_T^{\alpha^\star})-m(X_T^{\alpha^w})\right]/\sigma(X_T^{\alpha^\star})= 0.124$ that can be interpreted as the performance gain, measured in number of standard deviations, of the optimal strategy $\alpha^\star$ compared respectively to the constant strategy $\alpha^c$ and the WoMO strategy $\alpha^w$. Another interesting statistics is the surplus profit per trade $\left[ m(X_T^{\alpha^\star})-m(X_T^{\alpha^c})\right] / \left[ m(N^{b,\alpha^\star}_T)+m(N^{a,\alpha^\star}_T)+m(N^{market, \alpha^\star}_T) \right] = 0.056$ euros per trade, recalling that the maximum volume we trade is $\bar{\ell}=\bar{e}=100$. Note that for this last statistics, the profitable effects of the per share rebates $\rho$ are partially neutralized because the number of executions is comparable between $\alpha^\star$ and $\alpha^c$; therefore the surplus profit per trade is mainly due to the revenue obtained from \textit{making the spread}. To give a comparison point, typical clearing fee per execution is $0.03$ euros on multilateral trading facilities, therefore, in this backtest, the surplus profit per trade was roughly twice the clearing fees.

\begin{figure}[h!] 
\centering
\subfloat[N Bid empirical distribution]{\includegraphics[width=0.5\textwidth]{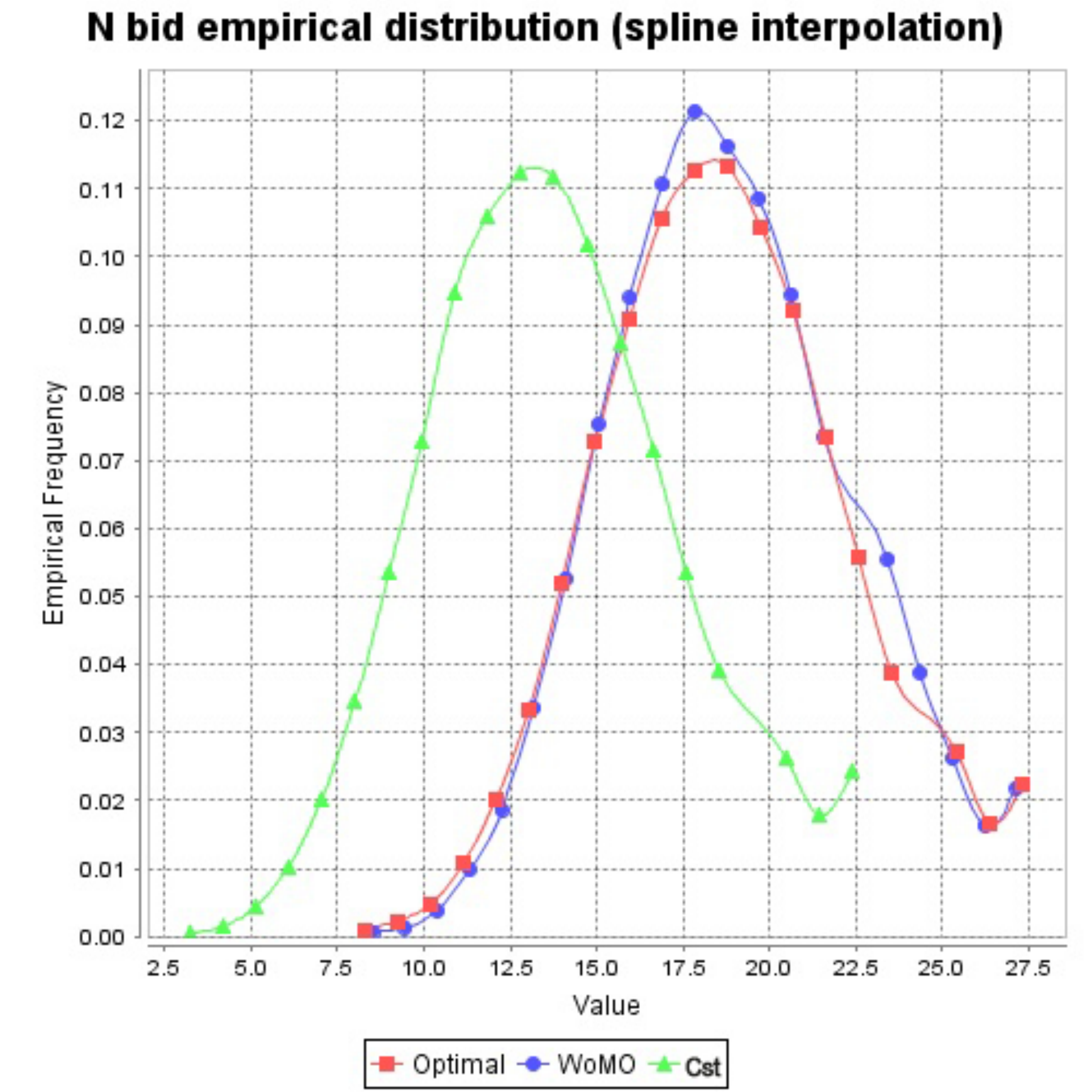}} 
\subfloat[N Ask empirical distribution]{\includegraphics[width=0.5\textwidth]{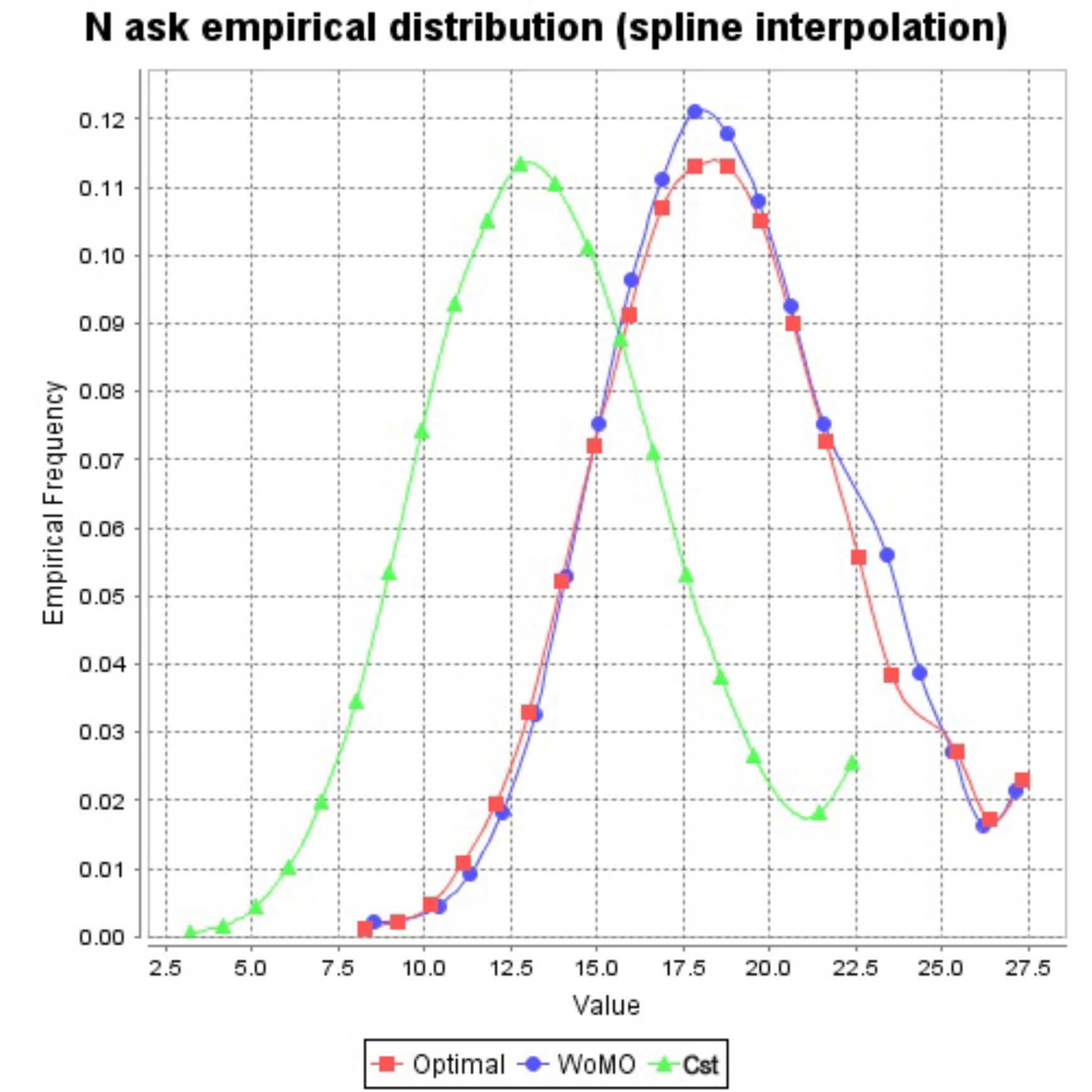}}
\caption{Empirical distribution of the number of executions on both sides.}
\label{empdisnumexec}
\end{figure}

We observe in the synthesis table that the number of executions at bid and ask are symmetric, which is also confirmed by the plots of their empirical distributions in figure \reff{empdisnumexec}. This is due to the symmetry in the execution intensities $\lambda^b$ and $\lambda^a$, which is reflected by the symmetry around $y=0$ in the optimal policy. 

Moreover, notice that the maximum absolute inventory is efficiently kept close to zero in $\alpha^\star$ and $\alpha^w$, whereas in $\alpha^c$ and $\alpha^r$ it can reach much higher values. The maximum absolute inventory is higher in the case of $\alpha^\star$ than in the case $\alpha^w$ due to the fact that $\alpha^\star$ can unwind any position immediately by using market orders, and therefore one may post higher volume for limit orders between two trading at market, profiting from reduced execution risk.

\vspace{2mm}

\begin{figure}[h!] 
\centering 
\includegraphics[width=0.7\textwidth]{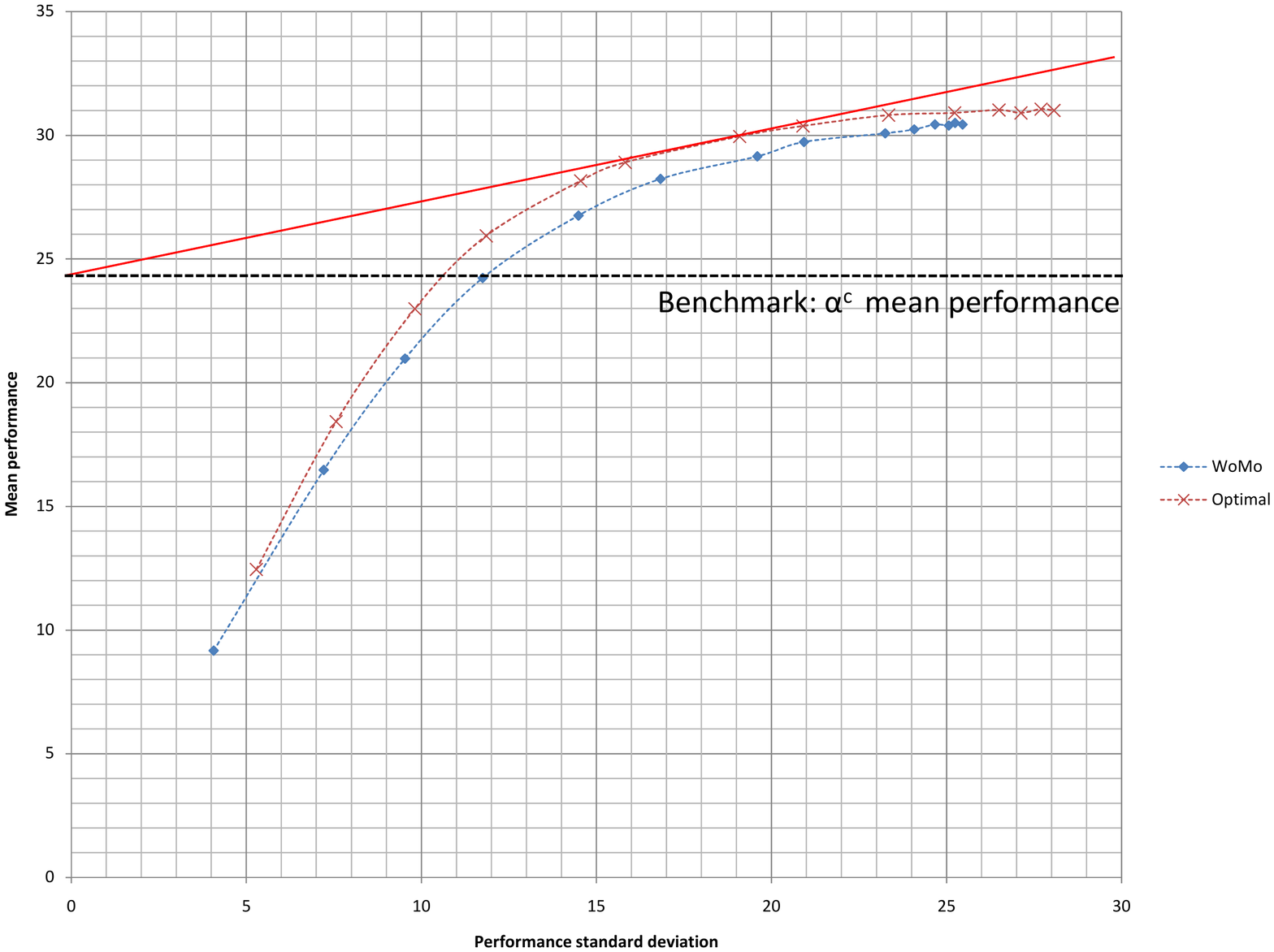} 
\caption{Efficient frontier plot}
\label{efffront}
\end{figure}

\noindent  {\bf  \em{Efficient frontier}.}
An important feature of our algorithm is that the market maker can  choose the inventory penalization parameter $\gamma$. To illustrate  its influence, we varied the inventory penalization $\gamma$ from $50$ to $6.10^{-2}$, and then  build the efficient frontier for both the optimal strategy $\alpha^\star$ and for the WoMO strategy $\alpha^w$. Numerical results are provided in table \reff{efficientfrontier} and a plot of this data is in figure \reff{efffront}.

\begin{table}[h!]
\footnotesize
\begin{center}
\begin{tabular}{|l	|	l	|	l	|	l	|	l	|	c	|   c |}
\hline 	$\gamma$ &	$\displaystyle \sigma(X_T^{\alpha^\star})$	&	$\displaystyle m(X_T^{\alpha^\star})$	&	$\displaystyle \sigma(X_T^{\alpha^w})$	&	$\displaystyle \sigma(X_T^{\alpha^w})$	&	$\operatorname{IR}(X^{\alpha^\star}_T)$ & $\operatorname{NIR}(X^{\alpha^\star}_T)$	\\
\hline 
50.000	&	5.283	&	12.448	&	4.064	&	9.165	&	2.356	&	-2.246	\\
25.000	&	7.562	&	18.421	&	7.210	&	16.466	&	2.436	&	-0.779	\\
12.500	&	9.812	&	22.984	&	9.531	&	20.971	&	2.343	&	-0.135	\\
6.250	&	11.852	&	25.932	&	11.749	&	24.232	&	2.188	&	0.136	\\
3.125	&	14.546	&	28.153	&	14.485	&	26.752	&	1.935	&	0.263	\\
1.563	&	15.819	&	28.901	&	16.830	&	28.234	&	1.827	&	0.289	\\
0.781	&	19.088	&	29.952	&	19.593	&	29.145	&	1.569	&	0.295	\\
0.391	&	20.898	&	30.372	&	20.927	&	29.728	&	1.453	&	0.289	\\
0.195	&	23.342	&	30.811	&	23.247	&	30.076	&	1.320	&	0.278	\\
0.098	&	25.232	&	30.901	&	24.075	&	30.236	&	1.225	&	0.261	\\
0.049	&	26.495	&	31.020	&	24.668	&	30.434	&	1.171	&	0.253	\\
0.024	&	27.124	&	30.901	&	25.060	&	30.393	&	1.139	&	0.242	\\
0.012	&	27.697	&	31.053	&	25.246	&	30.498	&	1.121	&	0.243	\\
0.006	&	28.065	&	30.998	&	25.457	&	30.434	&	1.105	&	0.238	\\
\hline 
\end{tabular}
\end{center}
\caption{Efficient frontier data}
\label{efficientfrontier}
\end{table}

We display both the ``gross" information ratio $\operatorname{IR}(X^{\alpha^\star}_T):= m(X^{\alpha^\star}_T)/\sigma(X^{\alpha^\star}_T)$ and the ``net" information ratio $\operatorname{NIR}(X^{\alpha^\star}_T):= \left( m(X^{\alpha^\star}_T)-m(X^{\alpha^c}_T)\right)/\sigma(X^{\alpha^\star}_T)$ to have more precise interpretation of the results. Indeed,  $m(X^\alpha_T)$ seems largely overestimated in this simulated data backtest compared to what would be real-world performance, for all $\alpha \in \lbrace \alpha^{\star},\alpha^{w},\alpha^{c},\alpha^{r} \rbrace$. Then, to ease interpretation, we assume that $\alpha^c$ has zero mean performance in real-world conditions, and therefore offset the mean performance $m(X^{\alpha^\star}_T)$ by the constant $-m(X^{\alpha^c}_T)$ when computing the NIR. This has simple visual interpretation as shown in figure \reff{efffront}.

Observe  that highest (net) information ratio is reached for $\gamma\simeq 0.8$ for this set of para\-meters. At this point $\gamma\simeq 0.8$, the annualized value of the NIR (obtained by simple extrapolation) is $47$, but this simulated data backtest must be completed by a backtest on real data.  Qualitatively speaking, the effect of increasing the inventory penalization parameter $\gamma$ is to increase the zone $\mathcal{T}$ where we trade at market. This induces smaller inventory risk, due to the fact that we unwind our position when reaching relatively small values for $\vert y \vert$. This feature can be used to enforce a soft maximum inventory constraint directly by choosing $\gamma$.

%\pagebreak 

\begin{small}

\end{small}

 \end{document}